\documentclass[
 reprint,
superscriptaddress,
 amsmath,amssymb,
 aps,
longbibliography
]{revtex4-2}

\usepackage{booktabs} 
\usepackage{graphicx}
\usepackage{array}[=2016-10-06]
\usepackage{dcolumn}
\usepackage{bm}
\usepackage{xcolor}
\usepackage{tikz}
\usepackage{lineno}
\usepackage{standalone}
\usepackage{siunitx}

\newcolumntype{C}[1]{>{\centering\arraybackslash}p{#1}}
\newcolumntype{H}[1]{>{\centering\arraybackslash}p{#1}} 
\newcolumntype{L}[1]{p{#1}} 

\begin{document}

\preprint{APS/123-QED}

\title{MAGNETO-$\nu$: Heavy neutral lepton search using $^{241}$Pu $\beta^-$ decays}

\author{C. Lee}
 \email{lee1118@llnl.gov}
\affiliation{Lawrence Livermore National Laboratory, 7000 East Ave, Livermore, CA 94550, USA}

\author{X. Zhang}
\affiliation{Lawrence Livermore National Laboratory, 7000 East Ave, Livermore, CA 94550, USA}
\author{A. Kavner}
\altaffiliation[Now at ]{University of Zurich, R\"amistrasse 71, 8006 Zürich, Switzerland}
\author{T. Parsons-Davis}
\affiliation{Lawrence Livermore National Laboratory, 7000 East Ave, Livermore, CA 94550, USA}
\author{D. Lee}
\affiliation{Lawrence Livermore National Laboratory, 7000 East Ave, Livermore, CA 94550, USA}
\author{N. Hines}
\affiliation{Lawrence Livermore National Laboratory, 7000 East Ave, Livermore, CA 94550, USA}

\author{S. T. P. Boyd}
\affiliation{
University of New Mexico, Albuquerque, NM 87131, USA
}%

\author{M. Loidl}
\author{X. Mougeot}
\author{M. Rodrigues}
\affiliation{%
Universit\'e Paris-Saclay, CEA, List, Laboratoire National Henri Becquerel (LNE-LNHB), 91120, Palaiseau, France}%

\author{M. K. Lee}
\author{J. W. Song}
\affiliation{%
Korea Research Institute of Standard and Science, 267 Gajeong-ro, Yuseong-gu, Daejeon 34113, Republic of Korea
}%

\author{R. Wood}
\author{I. Jovanovic}
\affiliation{%
University of Michigan, 2355 Bonisteel Blvd, Ann Arbor, MI 48109, USA
}%

\author{G. B. Kim}%
 \email{kim90@llnl.gov}
\affiliation{Lawrence Livermore National Laboratory, 7000 East Ave, Livermore, CA 94550, USA}

\date{\today}

\begin{abstract}
The MAGNETO-$\nu$ experiment searches for keV-scale heavy neutral leptons (HNLs) through precise measurements of the $\beta^-$-decay spectrum of $^{241}$Pu. We present spectra comprising a total of 194 million $\beta^-$ decays recorded using decay energy spectrometry with metallic magnetic calorimeters, representing the most statistically precise measurement of $^{241}$Pu $\beta^-$ decay to date. The $\beta$-endpoint energy was determined using $\gamma$ rays and X rays from an external $^{133}$Ba calibration source, yielding $Q_\beta = 22.273\,(33)$\,keV. The measured spectrum shows no statistically significant deviation from the allowed $\beta$-decay model. From a subset of the high-statistics data, we set an upper limit on the mixing of an 11.5-keV HNL with the electron neutrino, $|U_{e4}|^2 < 1.31 \times 10^{-3}$ at the 95\% confidence level.
\end{abstract}

\maketitle


\section{Introduction}
High-precision $\beta$-decay spectra provide critical insight into nuclear and particle physics, enabling measurements of the neutrino’s nonzero mass~\cite{katrin2022, echo2019, DeGerone2022980, PhysRevLett.114.162501, COSULICH1992143, DEPTUCK200080, deGroot_2023}, studies of axial-vector coupling ($g_A$) quenching in nuclear decays~\cite{PhysRevLett.133.122501, PhysRevC.110.055503}, investigations of the reactor-neutrino anomaly~\cite{PhysRevC.100.054323}, and searches for cosmic relic neutrinos~\cite{Betti_2019, deGroot_2023, PhysRevC.105.045501}. Furthermore, $\beta$-decay spectra enable searches for new physics beyond the Standard Model, including hypothetical heavy neutral leptons (HNLs) with right-handed chirality~\cite{PhysRevD.105.072004, HOLZSCHUH1999247, Hiddemann_1995,Troitsk_2017, ODragoun_1999,friedrich2021limits,martoff2021hunter} and possible scalar or tensor interactions in $\beta$ decays~\cite{PhysRevC.94.035503, PhysRevLett.125.112501}.

The MAGNETO-$\nu$ experiment searches for HNLs in the 1--20\,keV mass range by precisely measuring the $\beta^-$-decay spectrum of $^{241}$Pu. HNLs within this mass range are viable candidates for warm dark matter~\cite{10.1093/mnras/sty3057, Adhikari_2017}. With $Q_\beta \approx 22$\,keV and a decay spectrum closely resembling that of an allowed decay, $^{241}$Pu is a promising isotope for searches. The experiment employs decay energy spectrometry (DES)~\cite{LOIDL2008872, jang2012, app11094044} to measure nuclear-decay energy via temperature increases in the detector. DES offers a nearly unity detection efficiency for $\beta$ particles across a wide energy range, minimizing systematic uncertainties from dead layers and the need for efficiency corrections. Our DES system uses metallic magnetic calorimeters (MMCs) as a thermometer. MMCs are the most linear cryogenic sensors available, making them ideal for precision $\beta$-spectral shape measurements~\cite{Fleischmann2005, mmc_jltp_2018, PhysRevLett.125.142503}. This analysis is based on 194 million $^{241}$Pu $\beta^-$ decays—a statistical sample several hundred times larger than in previous measurements, representing the most precise $^{241}$Pu $\beta^-$ spectrum obtained to date.

This paper is organized as follows. Section II reviews the $^{241}$Pu $\beta^-$-decay theory; Section III describes the experimental setup and operation; Section IV outlines data processing and analysis; Section V presents the measured $^{241}$Pu $Q_\beta$ value; Section VI reports the $^{241}$Pu $\beta^-$-decay spectrum and corresponding HNL search based on the 194 million-decay dataset; and Section VII summarizes the conclusions.

\section{$\beta^-$-decay of $^{241}$P\MakeLowercase{u}}
\subsection{$\beta$ decay theory with heavy neutral leptons}
Fermi's theory of $\beta$ decay can be formulated as \cite{10.1119/1.1974382, PhysRevC.110.055503}:
\begin{eqnarray}\label{eq:beta_decay}
    N(E, m_\nu) dw = &&\frac{G^2_F \cos^2\Theta_c}{2 \pi^3} |M_{if}|^2  w\,  p\,  w_\nu\, \sum_{\nu} |U_{e\nu}|^2 p_\nu(m_\nu) \nonumber\\
    &&\times F(Z,w)\, X(w)\, r(Z,w)\, C(w).
\end{eqnarray}
Here $E$ denotes the $\beta$-particle kinetic energy, $m_\nu$ the neutrino mass eigenstate, $G_F$ the Fermi constant, $\Theta_c$ the Cabibbo angle, and $M_{if}$ the elements of the nuclear-transition matrix. The $\beta$-particle total energy and momentum are normalized to the electron rest mass, $m_e$, with $w = E/m_e + 1$ and $p = \sqrt{w^2-1}$. An antineutrino is emitted with a normalized total energy $w_\nu = w_0 - w$, where $w_0 = Q_\beta/m_e + 1$ defines the normalized endpoint energy. The neutrino mass $m_\nu$ modifies the normalized antineutrino momentum, $p_\nu = \sqrt{w_\nu^2 - (m_\nu/m_e)^2}$. The total rate $N(E, m_\nu) $ is the sum of the degenerate contributions of different $m_\nu$, each weighted by $|U_{e\nu}|^2$, where $U_{e\nu}$ are the elements of the Pontecorvo-Maki-Nagawa-Sakata (PMNS) matrix~\cite{10.1093/ptep/ptaa104}. The Fermi function $F(Z,w)$ accounts for the Coulomb interaction between the emitted $\beta$ particle and the daughter nucleus of charge $Z$~\cite{fermi_beta, RevModPhys.90.015008}. The term $X(w)$ corrects atomic-screening and exchange effects arising from the indistinguishability of electrons in quantum mechanics. 

Changes in atomic orbitals caused by nuclear decay also affect $E$; the corresponding atomic-overlap correction is given by~\cite{RevModPhys.90.015008}
\begin{equation}
    r(Z, w) = 1 - \frac{1}{w_0-w}\frac{\partial^2}{\partial Z^2} B(G).    
\end{equation}
Together, $X(w)$ and $r(Z,w)$ enhance the low-energy region of the spectrum while suppressing the high-energy tail. The shape factor $C(w)$ reflects nuclear-structure effects and is typically determined experimentally from deviations relative to allowed decays. For $^{241}$Pu, $C(w)$ is effectively unity and varies by no more than 0.3\% over the relevant energy range~\cite{Rizek1995}.

In the massless limit $m_\nu \to 0$, the antineutrino momentum $p_\nu$ converges to $w_\nu$, and Eq.~(\ref{eq:beta_decay}) can be rearranged as
\begin{equation}\label{eq:Kurie}
    \sqrt{\frac{N(E,0)}{p w F(Z,w)X(w)r(Z,w_0)}} \propto K (w_0 - w) \sqrt{C(w)}.
\end{equation}
A Kurie plot compares the left-hand side of Eq.~(\ref{eq:Kurie}) with $w$ (or equivalently $E$). For allowed $\beta$ decays, the Kurie plot forms a straight, descending line with a slope $K$. The $Q_\beta$ value is obtained from the intersection of the line with the energy axis.

\begin{figure}
\begin{tikzpicture}
\node (image) at (0,0) {
    \includegraphics[width=\linewidth]{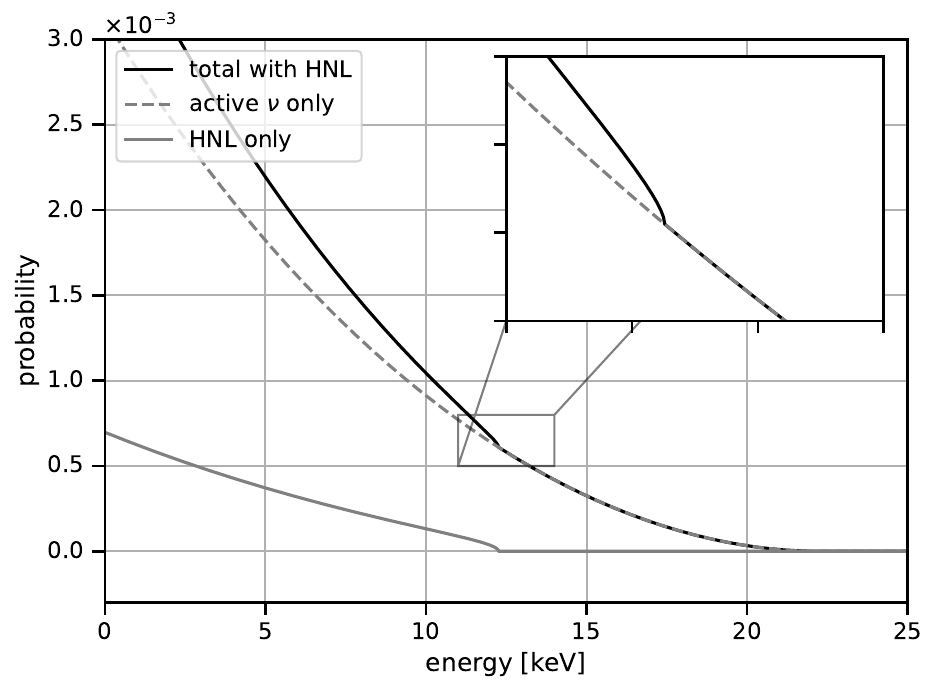}
    };
    \draw[thick, <->] (-2.2,-1.5) -- (-2.2,-.9);
    \node at (-2.3,-0.7) {\bf $|U_{e4}|^2$};
    \draw[thick, <->] (0.3,-2.) -- (3.1,-2.);
    \node at (2.,-2.2) {\bf $m_4$};
    \draw[thick, ->] (2.6, 2) -- (1.9,1.2) ;
    \node at (2.8, 2.2) {``kink"};
    \draw[thick, ->] (3.3, -1.5) -- (3.1,-1.85) ;
    \node at (3.5, -1.35) {$Q_\beta$};
\end{tikzpicture}
\caption{\label{fig:sterile_signal} HNL mixing signature in the $^{241}\text{Pu}$ $\beta$ decay spectrum.
Normal decays emitting active neutrinos terminate at $Q_\beta$ (dashed), whereas decays emitting hypothetical HNLs with mass $m_4$ = 10\,keV end at $Q_\beta - m_4$ (gray solid), near 12\,keV. The total spectrum (black solid), given by the sum of both decay modes, exhibits a characteristic “kink” at the endpoint of the HNL branch. An unusually large admixture, $|U_{e4}|^2 = 0.2$, is used here for illustration.} 
\end{figure}

A hypothetical HNL with a mass $m_4$ and an admixture $|U_{e4}|^2$ extends Eq.~(\ref{eq:beta_decay}), such that the total differential rate $R_\beta$ is
\begin{equation}\label{eq:sterile_rate}
R_\beta(E, m_4, |U_{e4}|^2) = (1-|U_{e4}|^2)N(E,m_\nu) + |U_{e4}|^2N(E,m_4).
\end{equation}
The first term represents the standard $\beta$-decay branch emitting active neutrinos, and the second term stands for the HNL-emitting branch. Figure~\ref{fig:sterile_signal} illustrates how a 10-keV HNL affects $R_\beta$. A hypothetical branch emitting an HNL has its endpoint shifted by $m_4$ (dotted line). This branch introduces a distinctive ``kink" at $Q_\beta - m_4$ in the combined differential spectrum (solid line). Such a kink should be detectable in a $\beta$ decay spectrum with high statistics and energy resolution.

\begin{figure}
\begin{tikzpicture}
\centering
\draw[ultra thick] (-1,2) -- (1,2);
\node[] at (0,1.7) {\textbf{$^{241}$Pu}};
\node[] at (-0.6,2.2) {\scriptsize 5/2+};
\draw[thick, dashed, ->] (1,2) -- (2,0.1);
\node[align=right] at (2.8,1) {$Q_{\beta} = 20.78$\,keV\\14.33 y};
\node[] at (0.9,0.5) {\scriptsize $\beta^-$: 99.998\%};

\draw[ultra thick] (2,0) -- (4, 0);
\node[align=center] at (3,-0.5) {432.6 y\\\textbf{$^{241}$Am}};
\node[] at (2.4,0.2) {\scriptsize $5/2-$};
\draw[dashed, ->] (2,0) -- (1,-4.4);
\draw[dashed, ->] (2,0) -- (1,-3.9);
\node[align=right] at (2.9,-3) {$Q_\alpha = 5637.82$\,keV};

\draw[ultra thick] (-1,-5) -- (1,-5);
\node[align=center] at (0,-5.5) {$2.14\times10^{6}$ y\\\textbf{$^{237}$Np}};
\node[] at (-0.6,-4.8) {\scriptsize 5/2+};
\node[] at (0.8,-4.8) {\scriptsize 0};

\draw[thick] (-1,-4.5) -- (1,-4.5);
\node[] at (0.6,-4.3) {\scriptsize 59.5};
\node[] at (-0.6,-4.3) {\scriptsize $5/2-$};
\node[] at (2.3,-1.7) {\scriptsize $\alpha$: 84.8\%};

\draw[thick] (-1,-4.) -- (1,-4.);
\node[] at (-0.6,-3.8) {\scriptsize $7/2-$ };
\node[] at (0.6,-3.8) {\scriptsize 103};
\node[] at (1.,-1.2) {\scriptsize $\alpha$: 13.1\%};

\draw[thick] (-1,-2.5) -- (1,-2.5);
\node[] at (-0.6,-2.3) {\scriptsize $3/2-$ };
\node[] at (0.6,-2.3) {\scriptsize 267.6};

\draw[thick] (-1,-2.) -- (1,-2.);
\node[] at (-0.6,-1.8) {\scriptsize$1/2-$ };
\node[] at (0.6,-1.8) {\scriptsize 281.4};

\draw[thick, ->] (0.2,-2.) -- (0.2,-2.5);
\draw[thick, ->] (0.2,-2.5) -- (0.2,-4.);
\node[] at (-1.8,-3) {\scriptsize $\gamma$ 164.6\,keV, 1.86\%};
\draw[thin] (-0.4,-3) -- (0.2,-2.9);
\draw[thick, ->] (0.,-2.5) -- (0.,-4.5);
\node[] at (-1.8,-3.3) {\scriptsize $\gamma$ 208.0\,keV, 21.2\%};
\draw[thin] (-0.4,-3.3) -- (0.,-3.2);
\draw[] (-0.4,-3) -- (0.2,-2.9);

\draw[thick, ->] (0.2,-4.) -- (0.2,-4.5);
\draw[thick, ->] (0.2,-4.5) -- (0.2,-5);
\draw[thick, ->] (-0.2,-4.) -- (-0.2,-5);

\node[] at (2.7,-4.4) {\scriptsize $\gamma$ 59.5\,keV, 35.9\%};
\draw[thin] (1.5,-4.5) -- (0.2,-4.8);

\draw[ultra thick] (-4,-1) -- (-2,-1);
\node[align=center] at (-3,-1.5) {6.75 d\\\textbf{$^{237}$U}};
\node[] at (-3.6,-0.8) {\scriptsize 1/2+};
\draw[dashed, ->] (-1,2) -- (-2,-0.9);
\node[align=right] at (-2.8,1.3) {$Q_\alpha = 5140.0$\,keV\\14.290 y};
\node[] at (-.5,1) {\scriptsize $\alpha$: 0.00245\%};

\draw[dashed, ->] (-2,-1) -- (-1,-1.9);
\node[] at (-1.,-1.3) {\scriptsize $\beta^-$: 51\%};

\draw[dashed, ->] (-2,-1) -- (-1,-2.4);
\node[] at (-2.0,-1.8) {\scriptsize $\beta^-$: 42\%};

\node[align=right] at (-2.7,-2.3) {$Q_{\beta} = 518.6$\,keV};

\end{tikzpicture}
\caption{Decay scheme of $^{241}$Pu \cite{ENDF-VII}. Most $^{241}$Pu nuclei undergo $\beta$ decay to $^{241}$Am with $Q_\beta$ = 20.78\,(17)\,keV~\cite{Wang_2021}. A small fraction undergoes $\alpha$ decay producing $^{237}$U, which subsequently $\beta$-decays.}\label{fig:pu241_decay_scheme}
\end{figure}

\subsection{$^{241}$Pu $\beta$ decay}
$^{241}$Pu $\beta$ decay is a first-forbidden, non-unique transition with a measured partial half-life of 14.329\,(29) years and a $Q_{\beta}=20.78\,(17)$\,keV~\cite{ENDF-VII, ODragoun_1999, Wang_2021}. The decay product $^{241}$Am undergoes $\alpha$ decay with a half-life of 432.6 (0.6) years, which is much longer than that of $^{241}$Pu~\cite{BASUNIA20062323}. Conversion and Auger electrons from $^{241}$Am decays constituted major backgrounds in earlier magnetic-spectrometer measurements~\cite{ODragoun_1999}. Besides the dominant $\beta$ branch, $^{241}$Pu undergoes a rare (0.00245 (1)\%) $\alpha$ decay to $^{237}$U with $Q_\alpha = 5140.0 (5)$\,keV. The daughter $^{237}$U subsequently emits $\beta$ particles.~\cite{BASUNIA20062323} ($t_{1/2}=6.752 (5)$ d, $Q_\beta = 518.6 (5)$\,keV). Figure~\ref{fig:pu241_decay_scheme} summarizes the decay scheme of $^{241}$Pu. 

High-precision measurements of the $^{241}$Pu $\beta$ spectrum have important implications for fundamental physics:
\begin{enumerate}
    \item The $Q_\beta \sim 20$\,keV of $^{241}$Pu is especially sensitive to $\sim$10\,keV HNLs that could constitute warm dark matter.
    
    \item  Because its $Q_\beta$ is among the lowest of known radionuclides, $^{241}$Pu is an excellent candidate for direct neutrino-mass measurement: spectral distortion near the endpoint is more pronounced and easier to detect. Background from $^{237}$U $\beta$ decay can be suppressed by identifying and rejecting its coincident $\gamma$ rays.

    \item $^{241}$Pu is also a promising target for cosmic-background-neutrino detection: its large capture cross section and low $Q_\beta$ increase sensitivity. Successful detection additionally requires identifying and rejecting the $^{237}$U $\beta$ background.

\end{enumerate}
A comparison with tritium is instructive. Both isotopes have $Q_{\beta}$ $\sim\mathcal{O}(20)$\,keV. Like tritium, $^{241}$Pu undergoes a single-branch $\beta$ decay without accompanying $\gamma$ or X rays. This simplicity reduces systematic uncertainties and simplifies data interpretation. However, $^{241}$Pu $\beta$ decay differs from tritium in two key ways: (i) The heavy-nucleus recoil is much less energetic, reducing resolution broadening compared to tritium; (ii) its decay energy arises mainly from rearrangement of atomic electrons rather than nucleon transitions~\cite{ODragoun_1999}. These differences make $^{241}$Pu-based experiments a complementary approach to tritium-based studies. At room temperature, $^{241}$Pu is easy to handle in its solid form, whereas tritium typically requires specialized facilities to contain the radioactive gas.
 
The precise $Q_\beta$ value of $^{241}$Pu remains uncertain. The currently accepted value derives largely from theoretical estimates~\cite{be2004monographie} and measurements of the mean decay energy~\cite{PhysRev.168.1398}. Loidl \emph{et al.} measured the $^{241}$Pu spectrum with MMC-based DES and reported a $Q_\beta$ value $\approx 0.8$\,keV above the accepted value~\cite{LOIDL20101454}. However, they did not claim a discovery because (i) no calibration source was available above $Q_\beta$, and (ii) their model underpredicted data below 3\,keV. Mougeot~\emph{et al.} later explained the low-energy excess by including atomic-screening and exchange effects~\cite{KOSSERT20111246, MOUGEOT2023111018}. The atomic overlap correction, which must be applied together with atomic corrections, implies that $Q_\beta$ is $\approx0.2$\,keV higher than the maximum kinetic energy of the $\beta$ particle.

We conducted our own calculation of $^{241}$Pu $\beta$-decay shape factor with the method described in \cite{KOSSERT2022110237, PhysRevC.107.024313, PhysRevC.110.055503}. Nuclear structure was determined with the NushellX code \cite{BROWN2014115} using the \textit{khpe} interaction in the \textit{jj67pn} valence space \cite{PhysRevC.43.602}. The latter was restricted to limit computational burden, disregarding the proton 1i$_{13/2}$ and neutron 1j$_{15/2}$ orbitals. The dominant matrix element corresponds to the zeroth multipole order, making the spectrum weakly sensitive to any variation of the effective value of the axial-vector coupling constant $g_A$. A linear shape factor similar to that of \cite{Rizek1995} was found, with a maximum difference of 0.7\%.

\section{Experimental Method}
\subsection{Decay Energy Spectrometry with Metallic Magnetic Calorimeter}
DES is a state-of-the-art technique that measures the total energy released in a radioactive decay using cryogenic microcalorimeters~\cite{app11094044}. It measures the temperature rise caused by the radioactive decays of sources embedded in a microscopic metallic absorber. The absorber, a thin gold foil, effectively stops charged decay products, including $\alpha$ and $\beta$ particles and daughter nuclei, and converts their kinetic energy to heat. It provides a powerful method for accurately determining the $\beta$-decay spectral shape because of its excellent energy resolution, nearly 100\% uniform detection efficiency, and the absence of a detector dead layer, which ensures a consistent response across the entire detection volume. DES is also unaffected by conversion and Auger electrons from $^{241}$Am decays, which dominated the background in the magnetic-spectrometer experiment~\cite{ODragoun_1999}. Electrons emitted in $^{241}$Am decay always accompany the $\alpha$ particle and therefore appear as high-energy events near 5.6\,MeV.

Among the microcalorimeter sensors for DES, an MMC is particularly well suited to high-precision $\beta$ spectrometry due to its excellent linearity and reproducibility~\cite{10.1063/1.4958699,mmc_jltp_2018,PhysRevLett.125.142503, kovavc2025comparison,u233_gamma}. An MMC converts deposited decay energy into a change in magnetization: energy deposition heats conduction electrons in the absorber, which alters the spin ordering of paramagnetic Er$^{3+}$ ions in the sensor material~\cite{enss2005low}. The signal conversion is highly linear over more than two orders of magnitude in energy, making MMCs ideal for mapping $\beta$-decay spectra across a broad range. 

The best energy resolution achieved with MMCs to date is 1.25\,eV full width at half maximum (FWHM) for 5.9\,keV photons~\cite{krantz2024magnetic}. The MMCs used in the MAGNETO-$\nu$ experiment exhibit lower resolution because their absorbers are several orders of magnitude larger to contain the high-activity source and fully stop $\alpha$ particles. Future detectors with smaller absorbers distributed over many multiplexed pixels could potentially reach eV-scale resolution \cite{echo_multiplexing,kempf2017demonstration}. 

MMCs also provide comparatively faster response times than other microcalorimeters of similar absorber size. This advantage stems from the fact that decay energy couples to the sensor solely through conduction electrons, which maintain relatively high thermal conductivity at millikelvin temperatures. The rapid response enhances timing resolution and event separation, allowing faster data collection for improved statistical precision.

\subsection{Source \& absorber preparation}\label{subsection:source}
The $^{241}$Pu source (Table \ref{tab:source}) was originally procured from Oak Ridge National Laboratory in 2009 and prepared at Lawrence Livermore National Laboratory (LLNL). On October 19, 2023, the source was purified using standard anion exchange chemistry to remove ingrown Am. Pu was dissolved in 8\,M nitric acid, and the solution was then loaded onto a column packed with Eichrom AG1-X4 resin. The column was sequentially washed with 8\,M nitric acid and 10\,M hydrochloric acid. Pu was eluted from the column with a 10\,M mixture of hydrochloric acid and concentrated hydroiodic acid (10:1 volumetric ratio). The purified Pu was converted into nitrate form by evaporating concentrated nitric acid several times, and the resulting nitrate was dissolved in 4\,M nitric acid. An aliquot was diluted to 0.09\,M nitric acid and deposited onto a 4\,mm × 6\,mm × 25\,\si{\micro\meter} gold foil using a micropipette for DES. The deposited solution, with a total activity of approximately 950\,Bq, was evaporated at $80^{\circ}\mathrm{C}$ on a hot plate. During the evaporation process, Pu-nitrate microcrystals could form, which temporarily absorb $\beta$-particle energy, distorting the $\beta$ spectrum, and degrading the energy resolution. To mitigate this, the gold foil was repeatedly folded and rolled with a jeweler’s mill to break down Pu-nitrate crystals. The folding and rolling process was repeated 20 times. This mechanical-alloying technique reduced the crystal size and improved the energy resolution~\cite{hoover2015measurement, kim2024decay}. Finally, the foil was encapsulated with an additional 5-\si{\micro\meter}-thick gold layer to prevent electron and $\alpha$-particle escape. The typical range of 20-keV $\beta$ particles in gold is $\approx1 \si{\micro\meter}$; thus a 5-$\si{\micro\meter}$-thick layer is sufficient to fully stop them. The thin gold layer also blocks $\mathcal{O}$(5)-MeV $\alpha$ particles from actinide decays. 

\begin{table}[b]
\caption{\label{tab:source}%
Isotopic compositions and relative activities of the MAGNETO-$\nu$ $^{241}$Pu source (decay-corrected from original certification to 2 Jan 2024)}
\begin{ruledtabular}
\begin{tabular}{cccccc}
&$^{238}$Pu & $^{239}$Pu & $^{240}$Pu & $^{241}$Pu & $^{242}$Pu \\
\colrule
atom \% & 0.08 & 6.30 & 23.96  & 54.70  & 14.96  \\
activity \% & 0.024 & 0.007 & 0.096  & 99.873  & 0.001  \\

\end{tabular}
\end{ruledtabular}
\end{table}


The gold foil was then divided into three pieces to reduce heat capacity and radioactivity. One of these pieces, labeled B-3, was then divided into two $\approx 1.6$\,mg pieces. The two pieces, named B-3.1 and B-3.2, were used for two detector setups, labeled Channel~0 and Channel~1. We measured $\alpha$ decays from the absorbers using an external $\alpha$ detector. The absence of detectable counts ($\ll 1$ cpm) confirmed negligible particle escape from the absorber, consistent with the Monte Carlo simulation presented below. The two samples were wire-bonded to MMC devices fabricated by the Korea Research Institute of Standards and Science (KRISS) in South Korea~\cite{song2024modification, song2025synthesis}. Table~\ref{tab:detectors} summarizes the two detector setups used in Run~122.

\begin{table}[b]
\caption{\label{tab:detectors}%
Run~122 of detector configurations}
\begin{ruledtabular}
\begin{tabular}{ccc}
& Channel 0  & Channel 1\\
\colrule
absorber mass (mg) &  1.6\,mg & 1.6\,mg \\
trigger rate & 90 cts/s & 90 cts/s \\
total counts & $89\times 10^6$ & $83\times 10^6$ \\
MMC chip &  KRISS 1 mm  &  KRISS 0.5 mm  \\
SQUID chip & \multicolumn{2}{c}{Magnicon XS1W \& X16W} \\
rise time & 0.45 ms &0.2 ms \\
decay time & 3.25 ms & 2 ms\\
signal size  & $0.8\, \Phi_0$/MeV& $0.6 \,\Phi_0$/MeV \\
\end{tabular}
\end{ruledtabular}
\end{table}

The Geant4 Monte Carlo framework~\cite{AGOSTINELLI2003250} was used to evaluate energy-loss mechanisms and their impact on the measured energy spectrum. Trajectories of $4\times10^6$ $\beta$ particles with kinetic energies between 0.5 and 25.5\,keV were simulated in a geometry closely matching the experimental absorber. The simulation employed Geant4's default \texttt{FTFP\_BERT\_HP} physics list, coupled with radioactive decays. 
Its standard electromagnetic physics was replaced with LLNL's low-energy electromagnetic models to accurately transport $\beta$ particles down to 10\,eV. This model also included bremsstrahlung via LLNL's Evaluated Electron Data Library (EEDL)~\cite{osti_5691165}. The simulation confirmed that no $\beta$ particles escaped the absorber. The main energy-loss mechanism was the escape of 9.9 and 11.8\,keV Au $L_\alpha$ and $L_\beta$ X rays. The total fractional energy loss was $\sim10^{-5}$ for 20\,keV $\beta$ particles, a negligible value that did not affect the measured spectrum.

\begin{figure*}
\begin{tikzpicture}
\node (image) at (0,0) {
    \includegraphics[width=0.6\linewidth]{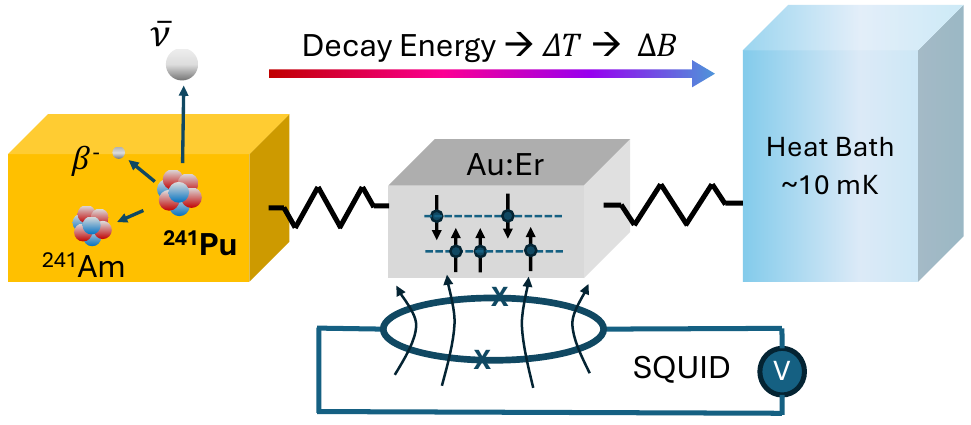}
    };
\node[align=left] at (-4.5,1.3) {\bf absorber};
\node at (-1.5,.5) {G};
\node[align=left] at (0.3,1.) {\bf MMC};
\node at (2,0.5) {G'};
\end{tikzpicture}
\begin{tabular}{cc}
    \centering

\includegraphics[width=0.4\linewidth]{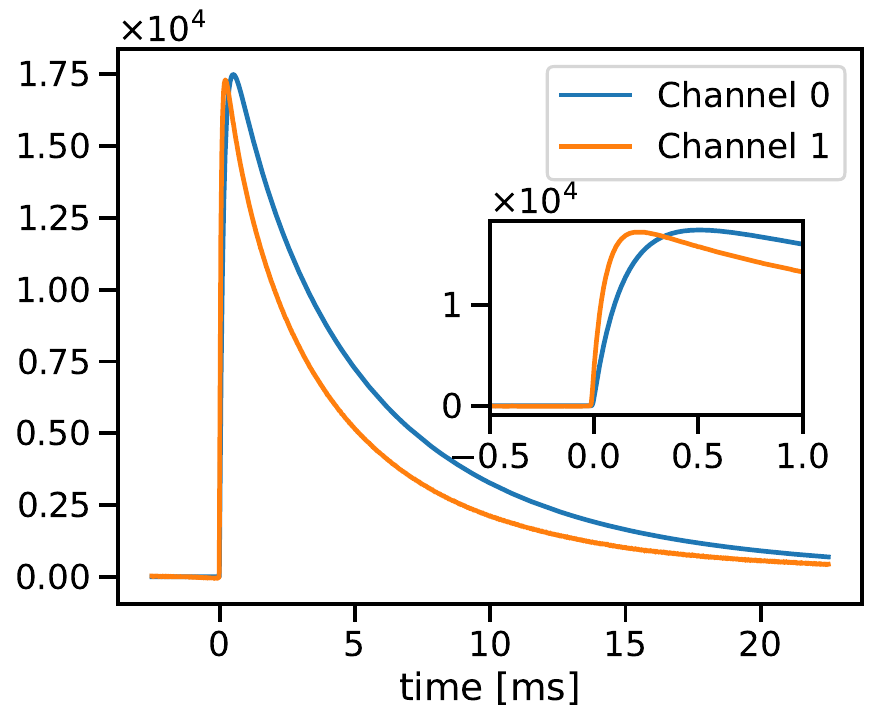}

\begin{tikzpicture}
\node (image) at (0,0) {
    \includegraphics[width=0.35\linewidth]{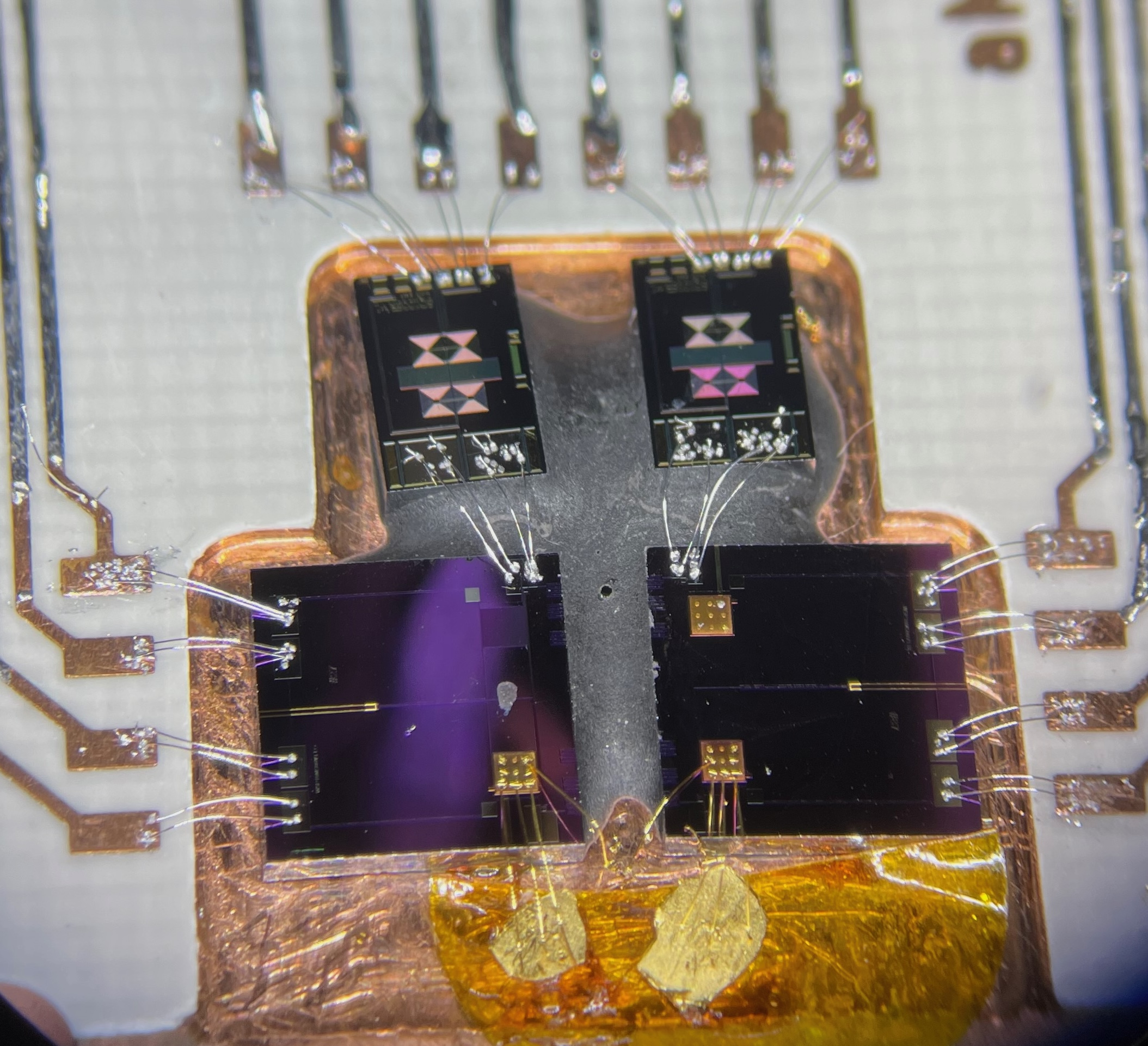}
    };
\node[] at (0.9,-3.2) {absorber};
\draw[thick, ->] (0.7,-3) -- (0.7,-2.4);
\node[] at (2.6,-3.2) {MMC};
\draw[thick, white, ->] (2.5,-3) -- (1,-1.4);
\node[thick] at (-2.2,0.9) {SQUID};
\node[align=right] at (-1,-3.2) {thermal bath};
\draw[thick, ->] (-1.,-3) -- (-1,-2.6);

\end{tikzpicture}
\end{tabular}

\caption{(Top) Principle of MMC-based DES. A $^{241}$Pu nucleus embedded in a gold absorber decays, and the decay energy is transferred to the Au:Er MMC sensor, altering its magnetization. (Left) Representative $\alpha$-decay signals from Run 122; \ Inset shows a zoomed view of pulse rise. Channel 1 pulses are faster due to smaller heat capacity of its MMC. The $y$-axis is in arbitrary units. (Right) Photograph of the MAGNETO-$\nu$ setup.} \label{fig:detector_setup}
\end{figure*}

\subsection{Experimental setup}
Figure~\ref{fig:detector_setup} shows a schematic of the MMC-based DES system, comprising a $^{241}$Pu-embedded absorber thermally coupled to an MMC sensor. The sensor was thermally linked to both the absorber and the thermal bath through 25-$\si{\micro\meter}$-diameter gold wires. An $\mathcal{O}(10)$-keV electron emitted in $^{241}$Pu decay raised the absorber temperature by approximately $\mathcal{O}(10)$\,\si{\micro\K} at 10\,mK. The thermalized energy was subsequently transferred to the Au:Er MMC sensor magnetized by a superconducting Nb coil. This transfer changed the sensor’s temperature and magnetization, and the SQUID detected the corresponding change in magnetization. The gold-wire bonds controlled heat flow and shaped the typical DES signals $V(t)$, which follow the form
\begin{equation}
    V(t) \propto e^{-t/t_r} - e^{-t/t_d}.
\end{equation}
Here, $t_r$ and $t_d$ denote the signal rise and decay times, respectively. The lower-left panel of Figure~\ref{fig:detector_setup} shows representative $\alpha$-decay pulses from the setup. The rise time depends on the absorber’s heat capacity relative to its thermal conductance to the MMC sensor, and the decay time depends on the same heat capacity relative to its conductance to the heat bath.

The lower-right panel of Figure~\ref{fig:detector_setup} shows the experimental assembly. The absorber, MMC sensor, and SQUID chips were mounted on a copper holder with GE-Varnish. The assembly was enclosed in a superconducting Nb can to shield the SQUIDs from stray magnetic fields. 

\subsection{Data acquisition}
The magnetic-flux signals from the MMC sensors were amplified by two SQUID stages in series, using XS1W and X16W units from Magnicon GmbH, Germany. The SQUIDs were operated in flux-locked-loop (FLL) mode using Magnicon's XXF-1 electronics. Most microcalorimeter experiments use the maximum FLL gain to maximize the signal amplitude relative to post-amplifier noise, thereby improving energy resolution. However, this high-gain setting limits the measurable energy range because the output voltage range of XXF-1 is only 10\,V. In this work, we prioritized a tenfold expansion of the measurable energy range at the cost of lower SQUID gain and correspondingly reduced energy resolution. This compromise enabled simultaneous acquisition of low-energy $^{241}$Pu $\beta$ and high-energy $\alpha$-decay signals. The $\alpha$-decay signals were later used to correct sensitivity drifts and to provide background-free energy calibration. The FLL output voltages were fed into SR-560 low-noise preamplifiers~\cite{stanford_research}, which also served as 100\,kHz low-pass filters. The filtered waveforms were continuously recorded at 200\,kS/s using a 14-bit NI PXIe-5172 oscilloscope unit~\cite{ni_5172} without hardware triggering to preserve complete data for offline software analysis.

The FLL electronics had limited bandwidth and occasionally lost lock during rapid flux changes following large $\alpha$ decays. Such losses produced sudden, discrete shifts in the FLL output voltage. These \emph{flux jumps} modified the FLL feedback current, inducing subtle changes in the DES unit’s temperature and sensitivity that required correction during analysis. The occurrence rate of SQUID \emph{flux jumps} varied from once every few minutes to once every few hours, depending on the specific SQUID unit, bias conditions, and input signal amplitudes.

This study focuses on two datasets collected during the first phase of the MAGNETO-$\nu$ experiment. 
The $^{241}$Pu $\beta$-spectrum shape and a preliminary HNL limit were derived from Run~122, which was acquired between December~25,~2023, and January~9,~2024, at 9\,mK. 
The $Q_\beta$ of $^{241}$Pu was measured in Run~151 (September~9–13,~2024) at 20\,mK using $\gamma$ rays and X rays from an external $^{133}$Ba calibration source.

\section{Data Processing \& Analysis}

\begin{figure}
\begin{tikzpicture}
\node (image) at (0,0) {
    \includegraphics[width=\linewidth]{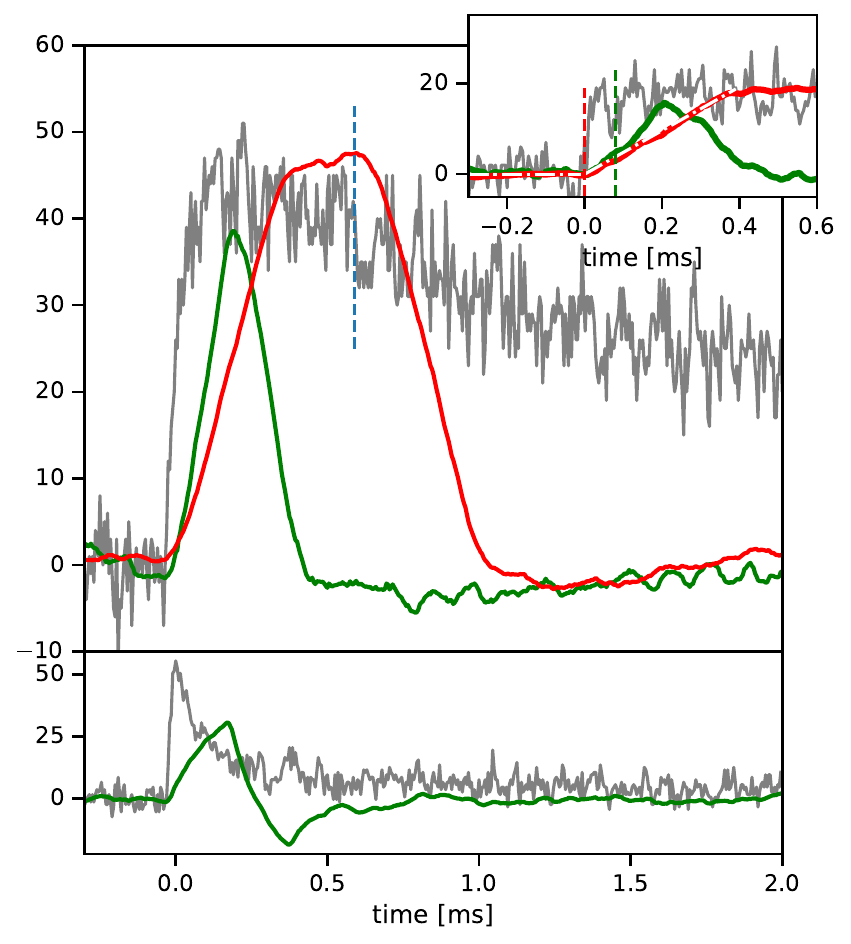}
    };
\node[align=left, green!30!gray] at (2.2,4.4) {\scriptsize Threshold crossing};
\node[align=left, red] at (1.2,4.2) {\scriptsize Pulse onset};
\node[align=left] at (2.6,3.2) {\scriptsize Fit};
\node[align=left] at (-2.2,4.) {Pulse amplitude};
\node[align=left, red] at (1.5,-0.6) {\emph{long pulse}};
\node[align=left, green!30!gray] at (-0.4,-.7) {\emph{short pulse}};
\draw[thick,-] (-1.4,3.8) -- (-0.8,3.3);
\node[align=left] at (-0.7,1.1) {$t_a$};

\node[align=left] at (2,0.2) {Absorber signal};
\node[align=left] at (2,-2.5) {Sensor-hit signal};

\node[align=left] at (0.9,-3.7) {Trigger minimum };
\draw[thick,-] (-1.3,-3.8) -- (-0.5,-3.7);
\end{tikzpicture}
\caption{(Top) Raw Channel 1 pulse (gray) with corresponding \emph{short}- (green) and \emph{long}- (red) trapezoid-shaped pulses. Pulse amplitude is measured at $t_a$ (blue dashed) on the \emph{long pulse}. (Inset) Zoomed-in view of a smaller pulse showing that the raw pulse begins (red dashed) 0.08 ms before the \emph{short pulse} crosses threshold (green dashed); $t = 0$ is defined by linear fits (white dot-dashed) to the \emph{long pulse}. (Bottom) A sensor-hit signal rises and decays faster than a typical absorber signal in the top panel. All $y$-axes are in arbitrary units.}
\label{fig:processing}
\end{figure}

The initial phase of MAGNETO-$\nu$ required collecting more than one billion $^{241}$Pu $\beta$ decays, with a target rate of 100\,Bq per detector, to achieve optimal sensitivity to $\mathcal{O}(10)$-keV HNLs. Recording an unbiased energy spectrum for both $\beta$ and $\alpha$ decays at such a high trigger rate is particularly challenging in microcalorimetry because of accidental coincidences. Rather than the optimal-filtering method commonly used in microcalorimeters, we employed trapezoidal shaping, which efficiently processes pulses shorter than a single decay time. Figure~\ref{fig:processing} shows a raw $\sim$18-keV pulse waveform and its two trapezoidally shaped versions: a \emph{short pulse} for triggering (green) and a \emph{long pulse} for amplitude measurement (red). These shaped pulses allow us to precisely determine the pulse amplitudes while minimizing bias from the trigger threshold and the amplitude dependence.

\subsection{Triggering \& efficiency}\label{subsection:trigger}
We employed a software trigger to detect pulses within the continuously recorded raw data stream. 
The \emph{short pulse} was shaped with a peaking time comparable to the signal rise time $t_r$~\cite{JORDANOV1994261}. 
Each trigger was followed by dead times of 1.25\,ms and 1\,ms for Channel~0 and Channel~1, respectively. 
A much longer 125\,ms dead time was applied after large pulses following $\alpha$ decays or SQUID \emph{flux jumps}. This was necessary because the undershoot of the \emph{short pulse} following such events would otherwise reduce the trigger efficiency for low-energy events. 
Because the DES detector provides nearly 100\% quantum efficiency for 20\,keV $\beta$ signals, dead-time losses were the dominant source of inefficiency.



\begin{figure}
\includegraphics[width=0.9\linewidth]{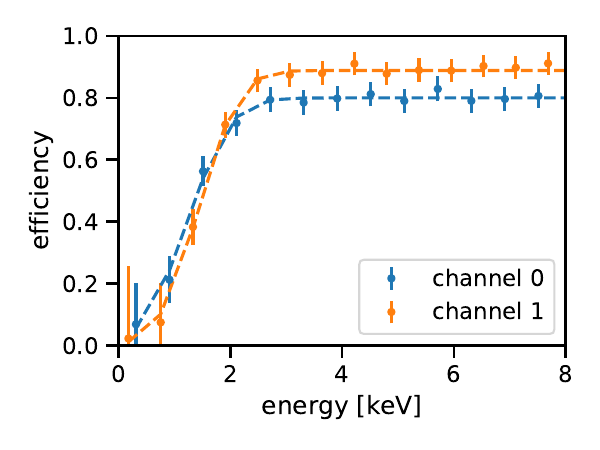}
\caption{Run~122 trigger efficiencies with best-fit sigmoid curves. Trigger dead time limited the asymptotic efficiency.}\label{fig:efficiency}
\end{figure}

The triggering efficiency was estimated by injecting synthetic pulses into the recorded data and counting how many were detected by the same software-trigger algorithm. Pulse templates were generated by taking the median of normalized, triggered signals. The templates were sufficiently long (125\,ms for Channel~0 and 75\,ms for Channel~1) to capture the full decay. A subset of raw data was selected, and simulated pulses with random amplitudes and timestamps were added at a rate of 1\,Hz. The resulting ``salted'' data were processed with the same trigger parameters as the original dataset. The efficiency was calculated as the fraction of injected signals successfully detected. Figure~\ref{fig:efficiency} shows the estimated efficiencies and their best-fit sigmoid curves. The centers of the best-fit sigmoid functions were 1.23\,keV and 1.43\,keV for Channel~0 and Channel~1, respectively. The efficiency saturated at 0.800 for Channel~0 and 0.888 for Channel~1 above 2\,keV, limited by the trigger dead time.

\subsection{Amplitude calculation}\label{subsection:amplitude}
We used the \emph{fixed-point-amplitude} method to extract pulse amplitudes. 
The method measured the \emph{long-pulse} value at a fixed time interval $t_a$ after the pulse began, minimizing bias in low-energy signals. In contrast, the conventional approach measures the maximum height of the trapezoid-shaped pulse, which tends to overestimate small-pulse amplitudes because noise adds a fixed width to the shaped waveform, a contribution that becomes proportionally larger for weaker signals.


To generate the \emph{long pulse}, we subtracted the median pre-trigger baseline from each triggered waveform and applied a longer peaking time to suppress high-frequency noise. 
The shaping also included a pole-zero (PZ) correction for the decay time $t_d$ and a flat-top interval comparable to $t_r$ to capture the full pulse height. 
The flat-time and PZ parameters were tuned to flatten the \emph{long pulse} near $t_a$, making the amplitude estimation robust against small jitters in determining the pulse start.

The \emph{fixed-point-amplitude} method required a precise and unbiased determination of the pulse start. 
The simplest raw timestamp, when the trigger pulse crossed the threshold, varied with pulse amplitude and trigger level. 
The inset of Figure~\ref{fig:processing} shows that the \emph{short pulse} crosses the threshold after the raw pulse had already started. 
This delay was more pronounced for smaller pulses, whose \emph{short-pulse} rise was slower, causing their amplitudes to be measured later than intended. 
To correct for this, we adopted the \emph{pulse-onset} method of~\cite{PhysRevD.111.052010} as the timing reference: 
the \emph{pulse onset} $t_0$ was defined as the intersection of a linear fit to the rising edge of the \emph{long pulse} with the mean pre-trigger baseline (white dot-dashed lines). 
This approach improved timing accuracy beyond the digitization limit by using the entire rising edge of the \emph{long pulse}. 
The pulse amplitude was extracted at $t_a$ after $t_0$, where $t_a$ equaled the peaking time plus 80\% of the flat-top duration. 
Figure~\ref{fig:processing} shows $t_a$ (0.59\,ms) located at the right edge of the flat-top region.

Baseline resolution was evaluated from noise traces processed identically to real pulses. The \emph{noise amplitudes} were centered near zero: $-0.100\,(0.348)$\,keV and $-0.073\,(0.387)$\,keV for Channel~0 and Channel~1, respectively. 
The standard deviations of these distributions define the baseline resolutions, corresponding to FWHM values of $0.820$\,keV for Channel~0 and $0.911$\,keV for Channel~1. 
Beyond the baseline, the DES energy resolution increased approximately linearly with energy, primarily due to residual errors from the \emph{correction} steps described in Section~\ref{subsection:correction}.

\subsection{Artifact removal}
\begin{figure}
\centering
\begin{tikzpicture}
\node (image) at (0,0) {
    \includegraphics[width=0.9\linewidth]{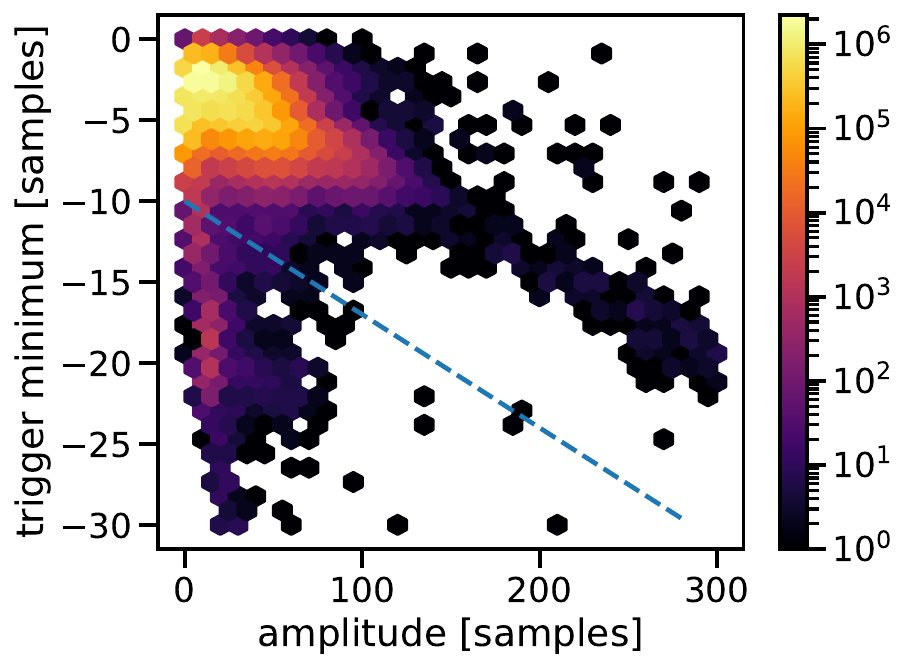}
    };
\node[] at (.5,-0.1) {Absorber};
\node[] at (-.3,-1.4) {Sensor hit};
\draw[thick, white, ->] (-1,-1.2) -- (-2,-0.5);
\end{tikzpicture}
\includegraphics[width=0.8\linewidth]{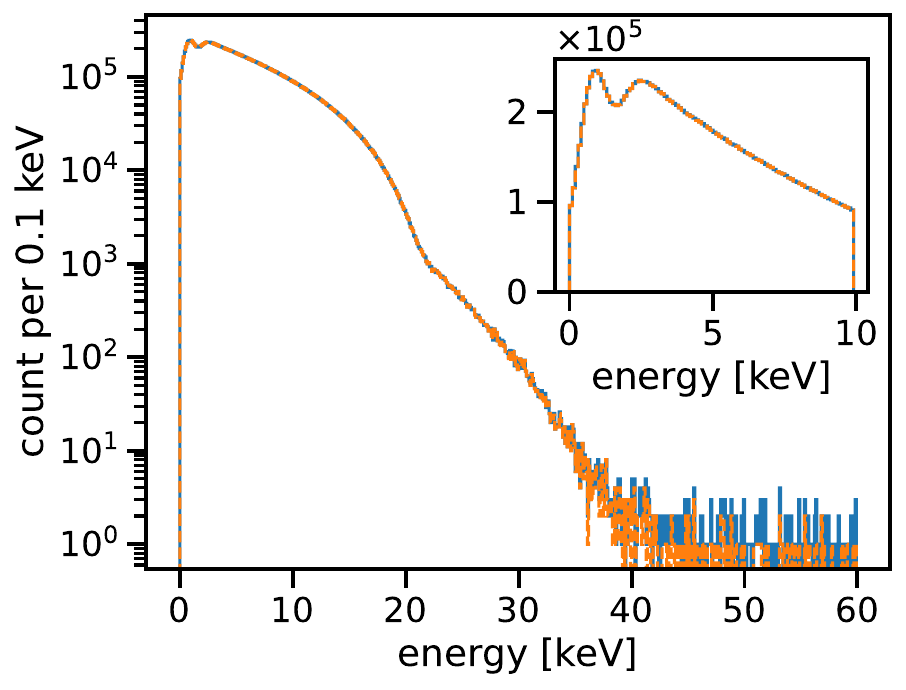}
\caption{(Top) Histogram of the \emph{long-pulse} amplitude ($x$-axis) versus the \emph{trigger minimum} ($y$-axis) for low-energy events in Channel~1. Normal absorber events appear above the dashed line, whereas sensor-hit events lie below and are excluded from further analysis. (Bottom) Energy spectra before (solid blue) and after (dashed orange) the cut are shown. Inset provides an expanded view of the spectrum below 10\,keV on a linear scale.}
\label{fig:dip_cut}
\end{figure}

The data contained several types of artifacts, which were rejected based on their distinct pulse shapes. These artifacts included (i) direct hits from external radiation on the MMC sensor, (ii) SQUID \emph{flux jumps}, and (iii) events occurring on the opposite side of the gradiometric MMC. Most of these artifacts exhibited rise and decay times shorter than 10\,\si{\micro\second}, and their \emph{short pulse} showed an abnormally large negative amplitude, as evident in the bottom panel of Figure~\ref{fig:processing}. 
The top panel of Figure~\ref{fig:dip_cut} compares these \emph{trigger minimums} with the \emph{amplitudes} of the \emph{long pulse}. For a given \emph{amplitude}, artifacts exhibited much larger \emph{trigger minimums} than normal pulses originating from the absorber. Events lying below the blue dashed line were therefore rejected. This cut removed fewer than $3\times10^{-6}$ of events at 4\,keV, a negligible fraction compared with the statistical uncertainty. Artifacts with small amplitudes could still pass the cut and contribute marginally to the dataset, so data below 4\,keV were excluded.

\begin{figure*}
\begin{tabular}{cc}
    \centering
        (a) amplitude over time & (b) baseline over time \\[6pt]
    \includegraphics[width=0.4\linewidth]{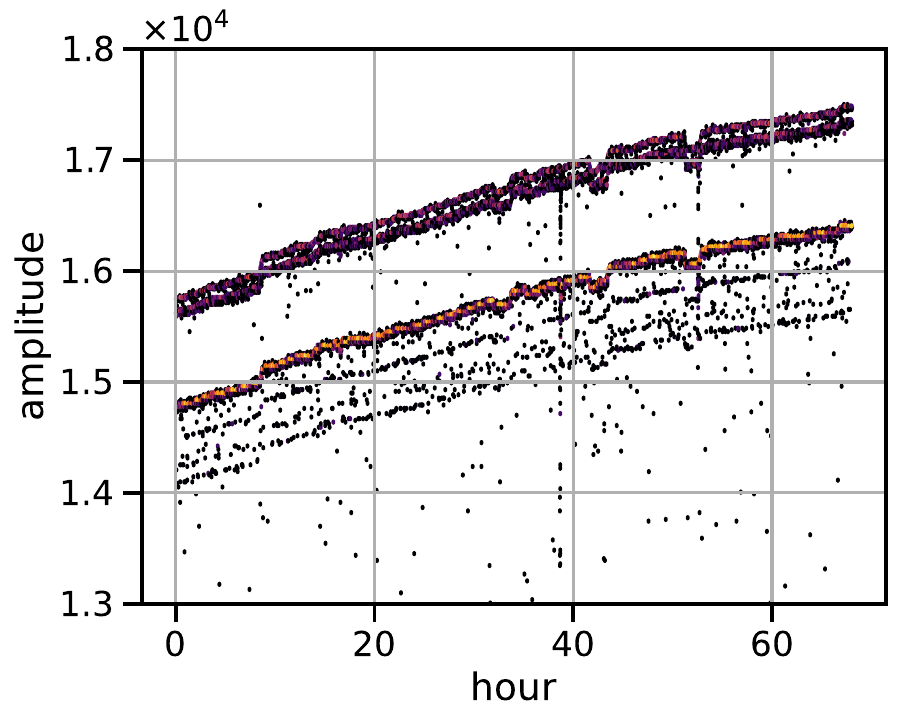} & \includegraphics[width=0.4\linewidth]{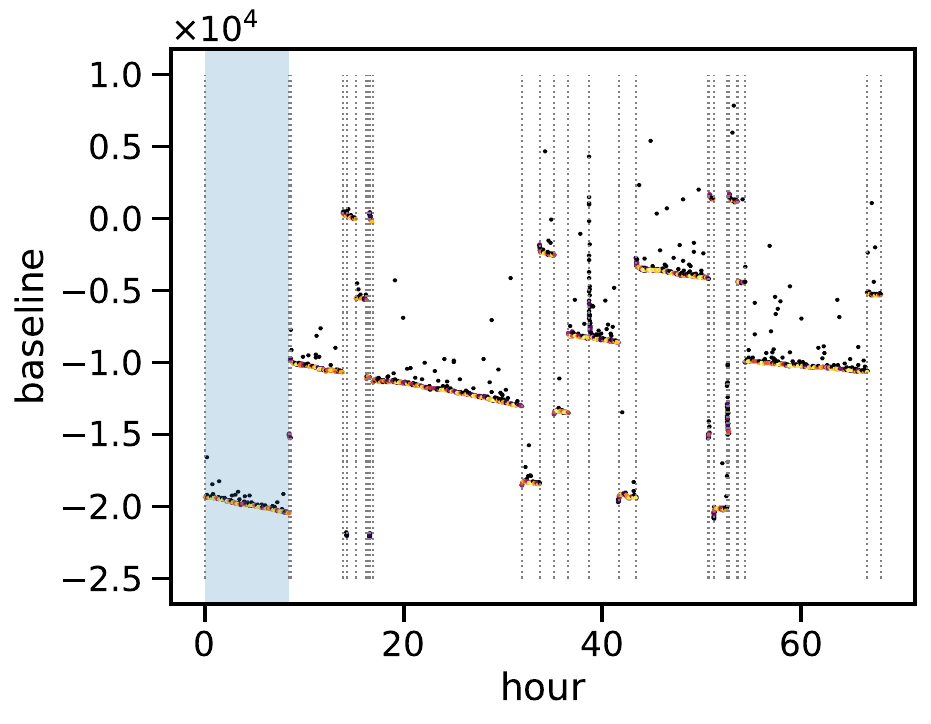} \\
    (c) amplitude vs. baseline & (d) corrected amplitude \\[6pt]
    \includegraphics[width=0.4\linewidth]{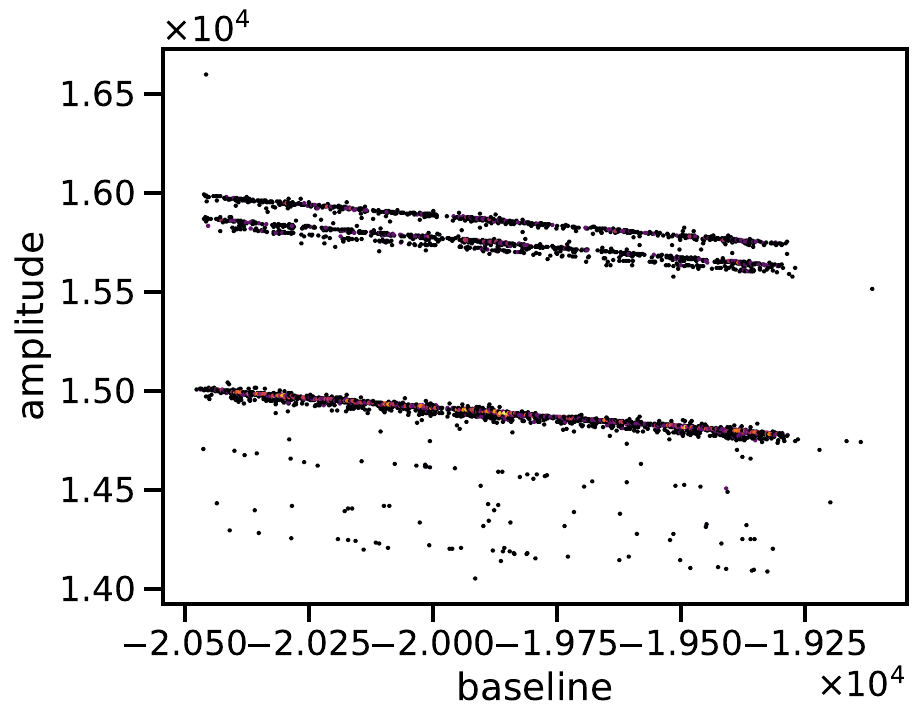} &    \includegraphics[width=0.4\linewidth]{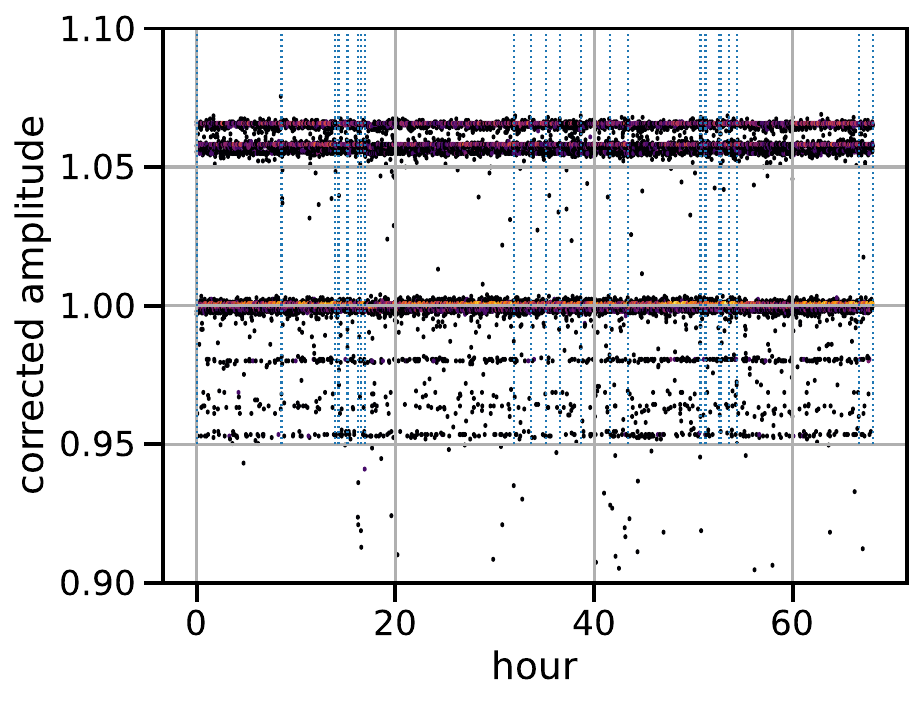} \\
    \end{tabular}
\caption{\emph{Baseline-correction} procedure.
(a) $\alpha$-decay amplitudes increase over time as the detector’s sensitivity rises.
(b) The pre-trigger baseline decreases over time. SQUID \emph{flux jumps} (vertical dotted lines) produce spontaneous baseline shifts.
(c) An anti-correlation between the $\alpha$-decay amplitude and the baseline in the first time interval (blue-shaded region in (b)) is measured and used to correct the sensitivity drift of all nearby events.
(d) This correction is repeated for each time interval, yielding $\alpha$ amplitudes with reduced scatter over time.}
    \label{fig:baseline_correction}
\end{figure*}

\subsection{Sensitivity drift corrections}\label{subsection:correction}
The primary source of systematic error in our system was time-dependent variation in detector sensitivity. This variation arose because the detector temperature drifted over time, and the DES signal amplitude is inversely proportional to temperature: higher temperatures increase the heat capacity and reduce the signal size. To correct for this effect, we measured the correlation between the signal baseline, used as a proxy for the MMC sensor temperature, and the $\alpha$-decay signal amplitudes. This correlation was then applied to correct the sensitivity drifts in all nearby events.
Figure~\ref{fig:baseline_correction} summarizes the detailed steps of this procedure, referred to as the \emph{baseline correction}. 
After the initial \emph{baseline correction}, a smaller residual fluctuation remained. 
It was characterized using a rolling median of 200 $^{241}$Am $\alpha$ decays at 5637.8\,keV and applied to correct the sensitivity drift in nearby events.
The correction sequence was repeated for each interval between SQUID \emph{flux jumps} to account for subtle variations in detector conditions following each \emph{flux jump}.

\begin{figure}
    \begin{tikzpicture}
    \centering
    \node (image) at (0,0) {\includegraphics[width=0.9\linewidth]{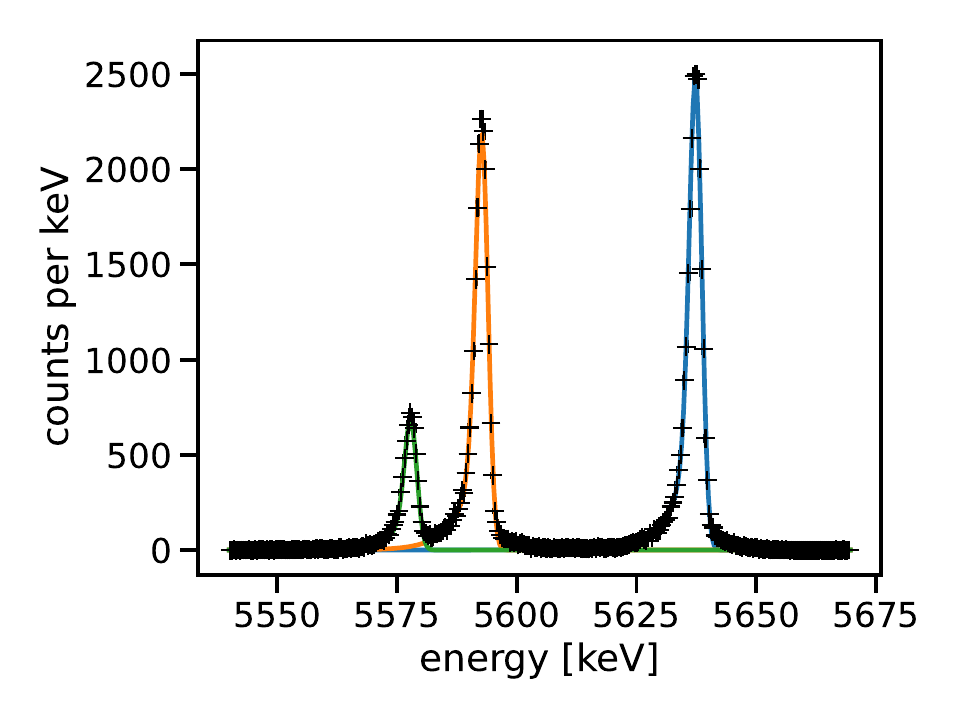}
    };    
    \node[align=left] at (2.5,1.5) {$^{241}$Am};
    \node[align=left] at (-0.6,2) {$^{238}$Pu };
    \node[align=left] at (-1.2,0.3) {$^{241}$Am \\- 59.5\,keV $\gamma$};
    \end{tikzpicture}
    \caption{$^{241}$Am (5637.82\,keV) and $^{238}$Pu (5593.27\,keV) peaks from Channel 1 with fitted EMG functions. $^{241}$Am $\alpha$ decays also produce a 5578.28 keV peak when a 59.5 keV $\gamma$ ray escapes the absorber without depositing its full energy.}
    \label{fig:am241_peaks_fit}
\end{figure}

\subsection{Energy calibration}\label{subsec:calibration}
A major challenge in energy calibration for DES-based precision $\beta$ spectrometry is Compton scattering from the calibration photons. While the DES absorbers are sufficiently thick to stop charged particles, they cannot fully absorb photons with energies comparable to the $\beta$ spectrum. As a result, Compton scattering inevitably introduces additional counts and distorts the spectral shape. Although one can alternate external photon calibration sources on and off, ensuring calibration consistency and correcting sensitivity drifts during the off periods, when no monoenergetic reference lines are available, remains challenging.
To address this issue, the energy calibration for Run~121 used monoenergetic actinide $\alpha$ decays from $^{238}$Pu, $^{239}$Pu, $^{240}$Pu, $^{242}$Pu, and $^{241}$Am. 
As described in Section~\ref{subsection:source}, these isotopes were part of the existing $^{241}$Pu source in the absorber, and their decays did not produce signals within the $^{241}$Pu $\beta$-decay region. 

The amplitudes of these $\alpha$ decays were extracted from the peak centroids of the intrinsic detector response modeled with an exponentially modified Gaussian (EMG). Because energy deposited in the source crystal thermalizes slowly, a portion does not reach the gold absorber within the signal window, causing the measured energies to fall below the true decay energy. To model this behavior consistently, a composite EMG function with one Gaussian width and two exponential decay constants was applied to all $\alpha$ peaks. Figure~\ref{fig:am241_peaks_fit} shows example $^{241}$Am and $^{238}$Pu $\alpha$-decay peaks accurately fitted with the composite-EMG detector response function.
Low-energy $\beta$ decays could also experience similar signal deformation. We estimate the maximum energy loss of 20-keV $\beta$ particles to be 24\,eV, based on a comparison of the stopping powers of 20-keV $\beta$ particles and 5-MeV $\alpha$ particles in a plutonium matrix, as reported in the NIST database~\cite{nist_star}.

\begin{figure*}
\begin{tabular}{cc}
    \centering
    \begin{tikzpicture}
    \node (image) at (0,0) {
    \includegraphics[width=0.5\linewidth]{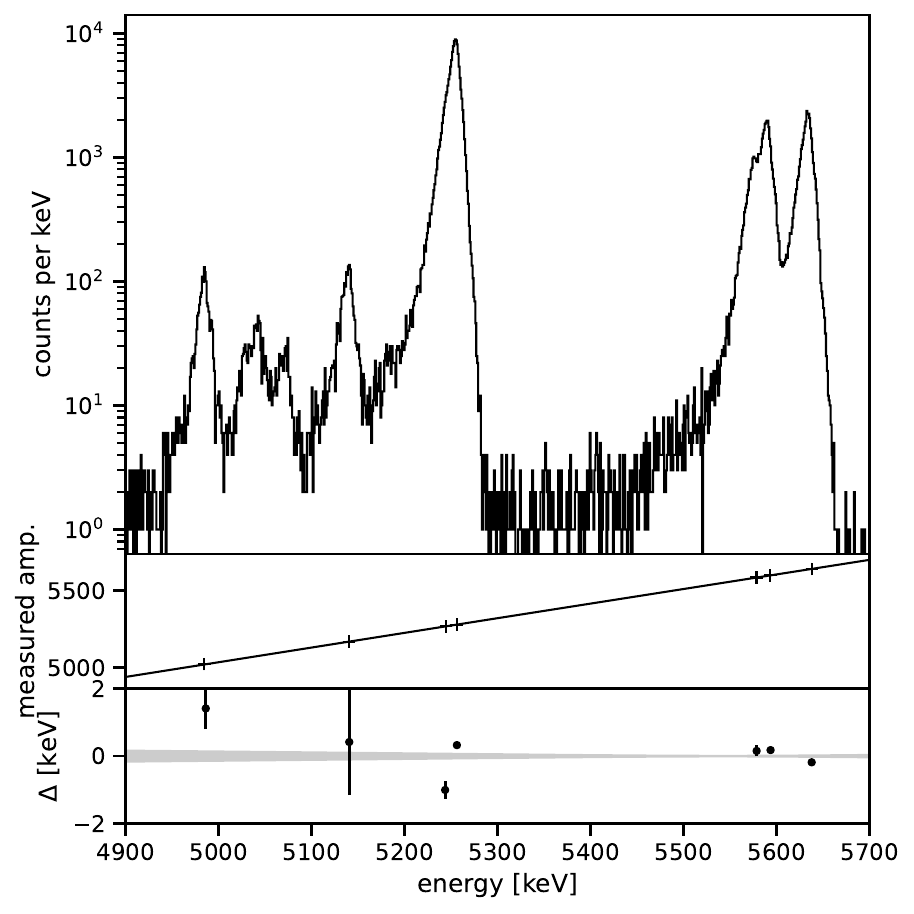}
    };
    \end{tikzpicture}
     & \begin{tikzpicture}
    \node (image) at (-0.2,-1.4) {
    \includegraphics[width=0.5\linewidth]{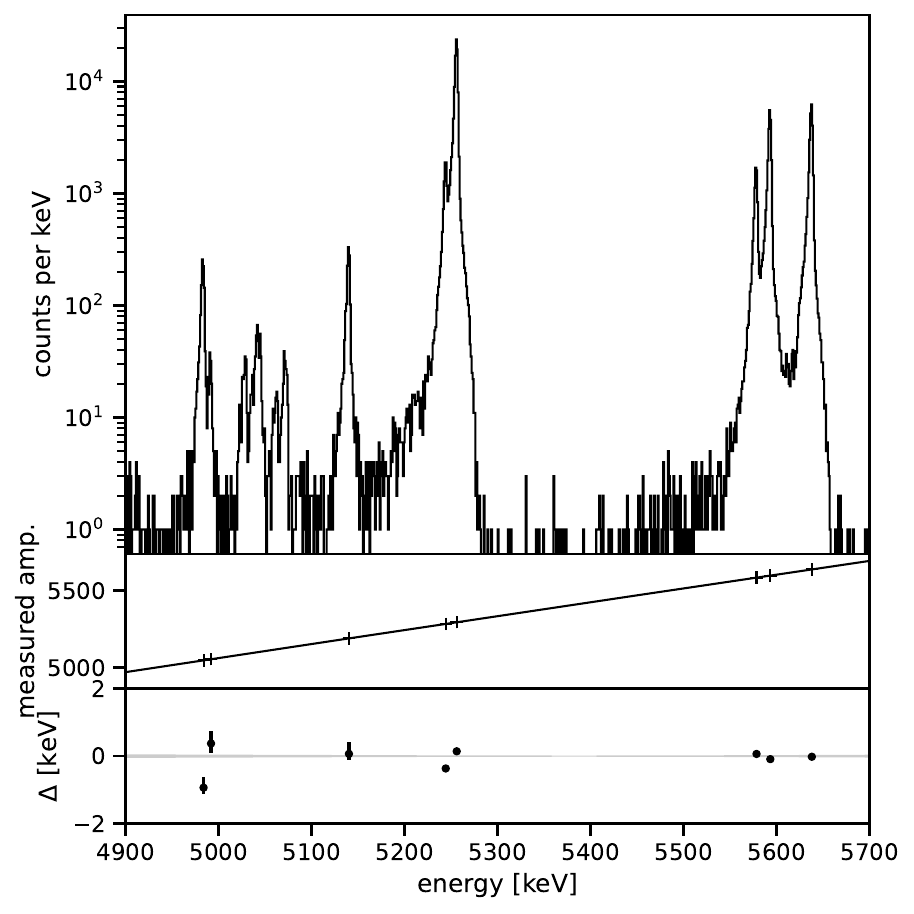}
    };
    \node[] at (-2.6,1.2) {$^{242}$Pu};
    \node[align=right, font=\linespread{0.8}\selectfont] at (-2, 0.5) {$^{241}$Pu $\alpha$\\ - X-rays};
    \node[] at (-1.3,1.1) {$^{241}$Pu $\alpha$};
    \node[] at (-1.2,2.1) {$^{239}$Pu};
    \draw[thick, ->] (-0.8,1.9) -- (-0.4,1.5);
    \node[] at (0.3, 2.2) {$^{240}$Pu};
    \node[align=right] at (1.7, 1.4) {$^{241}$Am - $\gamma$};
    \node[] at (2.2,2) {$^{238}$Pu};
    \node[] at (3, 2.2) {$^{241}$Am};
    \end{tikzpicture}
\end{tabular}
\caption{(Top) $\alpha$-decay peaks in Channel 0 (left) and Channel 1 (right). (Middle) Measured $\alpha$-peak centroids with literature values. Solid lines show the best-fit detector responses. (Bottom) Calibration residuals in keV with 1-$\sigma$ error bars. Gray bands represent calibration uncertainties. Discrepancies in $^{239}$Pu and $^{240}$Pu peak energies are likely due to errors in literature values, consistent with patterns observed in other independent MMC-based DES spectra (not shown).}
\label{fig:alpha_peaks}
\end{figure*}

We used a quadratic calibration function to describe the MMC response:
\begin{equation}
E = p_1 (e-e_0) + p_2 (e-e_0)^2. \label{eq:quadratic}
\end{equation}
Here, $e$ denotes the measured amplitude and $E$ the corresponding true energy in keV. The parameter $p_1$ denotes the gain, and $p_2$ characterizes the nonlinearity, primarily arising from the temperature-dependent susceptibility of MMC. The term $e_0$ corresponds to the \emph{noise amplitude} ($E = 0$) distribution defined in Section \ref{subsection:amplitude} and served as the low-energy calibration point. The values of $p_1$ and $p_2$ were obtained by comparing the measured $\alpha$-peak centroids with literature energies. Equation~(\ref{eq:quadratic}), using the best-fit parameters, was applied to convert measured amplitudes to true energies. The detector responses remained highly linear. Nonlinearity contributed 0.022\% (Channel~0) and 0.050\% (Channel~1) at 22\,keV, increasing to 5.0\% and 11.4\% at 5\,MeV.

We note that this calibration had a limitation at the $\beta$ energies of interest because most reference peaks lay at much higher energies of $\mathcal{O}(5)$\,MeV. Although MMCs exhibits excellent linearity, the lack of calibration points near the $\beta$ region could introduce systematic uncertainties from higher-order nonlinearities beyond the quadratic term that could not be constrained by the $\alpha$-decay peaks or the \emph{noise amplitudes} alone. Because the HNL search is primarily sensitive to spectral-shape distortions, uncertainties in the absolute energy scale play only a minor role. Nevertheless, we rescaled the calibrated spectrum using the independently measured $Q_\beta$ to ensure accurate comparison with theoretical beta-decay spectra and HNL models.


\subsection{$\alpha$ decay spectra}
Figure~\ref{fig:alpha_peaks} presents the final $\alpha$ spectra from Run~122. Channel 0 and Channel 1 contain $2.2\times10^5$ and $1.7\times10^5$ $\alpha$ decays, representing an unprecedented statistical sample. The peaks correspond to $\alpha$ decays of $^{242}$Pu, $^{241}$Pu, $^{240}$Pu, $^{239}$Pu, $^{241}$Am with 59.54-keV $\gamma$-ray escape, $^{238}$Pu, and $^{241}$Am. The middle panels compare the measured peak amplitudes with literature energies. The calibration curves between these values are nearly linear. The lower panels display the calibration residuals. The error bars represent combined uncertainties from centroid determination and literature energies. The gray bands indicate detector calibration uncertainties. All peaks align within 2\,keV for Channel~0 and 1\,keV for Channel~1. The FWHM values of the $\alpha$ peaks are 10.8\,keV (Channel 0) and 2.9\,keV (Channel 1). Channel~0 did not clearly resolve peak doublets such as $^{239}$Pu and $^{240}$Pu because of its inferior energy resolution. This broader resolution is attributable to slower signal rise times, more frequent SQUID \emph{flux jumps}, every few minutes instead of hours, and less precise baseline-dependent sensitivity corrections. These combined effects lead to a larger calibration uncertainty for Channel 0.

\begin{figure}
\begin{tikzpicture}
\node (image) at (0,0) {
    \includegraphics[width=\linewidth]{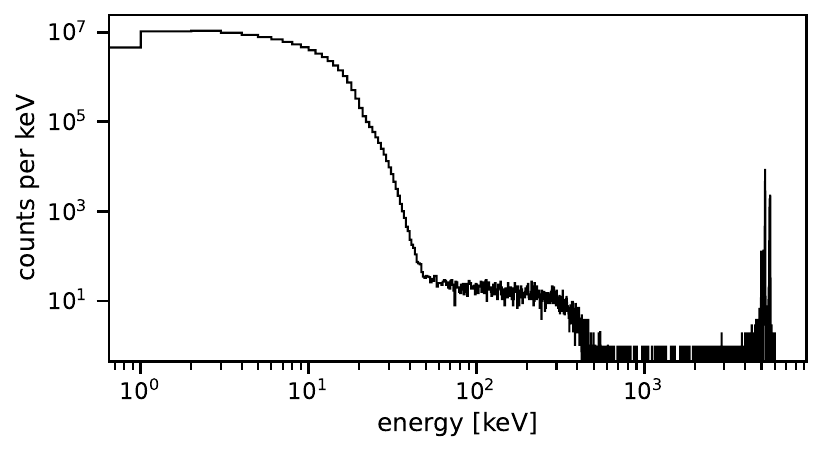}
    };
\node[] at (-1.4,1.3) {$^{241}$Pu $\beta$};
\node[] at (1.5,1.7) {$^{241}$Pu $\beta$ coincidences};
\draw[gray, thick, ->] (0.2,1.5) -- (-0.1,0.5);
\node[] at (1, -0.2) {$^{237}$U $\beta$};
\node[align=right] at (2.7,1.) {Am \& Pu $\alpha$'s};
\end{tikzpicture}
    
\begin{tikzpicture}
\node (image) at (0,0) {
    \includegraphics[width=\linewidth]{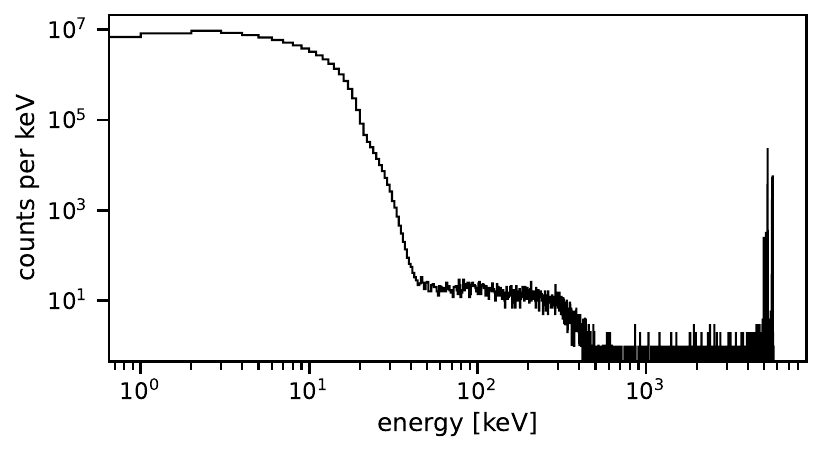}
    };
\node[] at (-1.4,1.3) {$^{241}$Pu $\beta$};
\node[] at (1.5,1.7) {$^{241}$Pu $\beta$ coincidences};
\draw[gray, thick, ->] (0.2,1.5) -- (-0.2,0.5);
\node[] at (1, -0.2) {$^{237}$U $\beta$};
\node[align=right] at (2.7,1.) {Am \& Pu $\alpha$'s};
\end{tikzpicture}

\caption{\label{fig:ch1_global_spectrum} Total energy spectra from Run 122 for Channel 0 (top) and Channel 1 (bottom).}
\end{figure}

Figure~\ref{fig:ch1_global_spectrum} displays the total energy spectrum from Run~122 for both channels. At low energies, most events originated from $^{241}$Pu $\beta$ decays, and accidental coincidences produce a broad hump above 20\,keV. The flat background from 40 to 518.6\,keV arose mainly from $^{237}$U $\beta$ decays. The peaks around 5\,MeV correspond to $\alpha$ decays, as discussed in Figure~\ref{fig:alpha_peaks}.

\section{$^{241}$P\MakeLowercase{u} $Q_\beta$ measurement}\label{PuQ}

The $Q_\beta$ value of $^{241}$Pu was determined in an independent experimental run, Run~151, using an external $^{133}$Ba calibration source. Positioned outside the cryostat, the source irradiated the detector with 30.97313\,(46)\,keV and 30.62540\,(45)\,keV $^{133}$Cs $K_{\alpha1}$ and $K_{\alpha2}$ X rays through thin aluminum windows~\cite{RevModPhys.75.35}. Below the $^{241}$Pu $Q_\beta$, the same source’s 80.9979\,(11)\,keV $\gamma$ ray produced escape peaks at 12.191 and 14.005\,keV, corresponding to the escape of 68.8069\,(22)\,keV and 66.9930\,(23)\,keV Au $K_{\alpha1}$ and $K_{\alpha2}$ fluorescence X rays from the absorber. Interpolation between these peaks provided a precise determination of $Q_\beta$. To isolate the intrinsic $\beta$ spectrum, measurements were performed both with and without the $\gamma$-ray source under identical conditions in the same run, enabling a direct comparison and a $Q_\beta$ determination free from $\gamma$-induced effects.

\begin{figure}
\begin{tikzpicture}
\node (image) at (0,0) {
    \includegraphics[width=\linewidth]{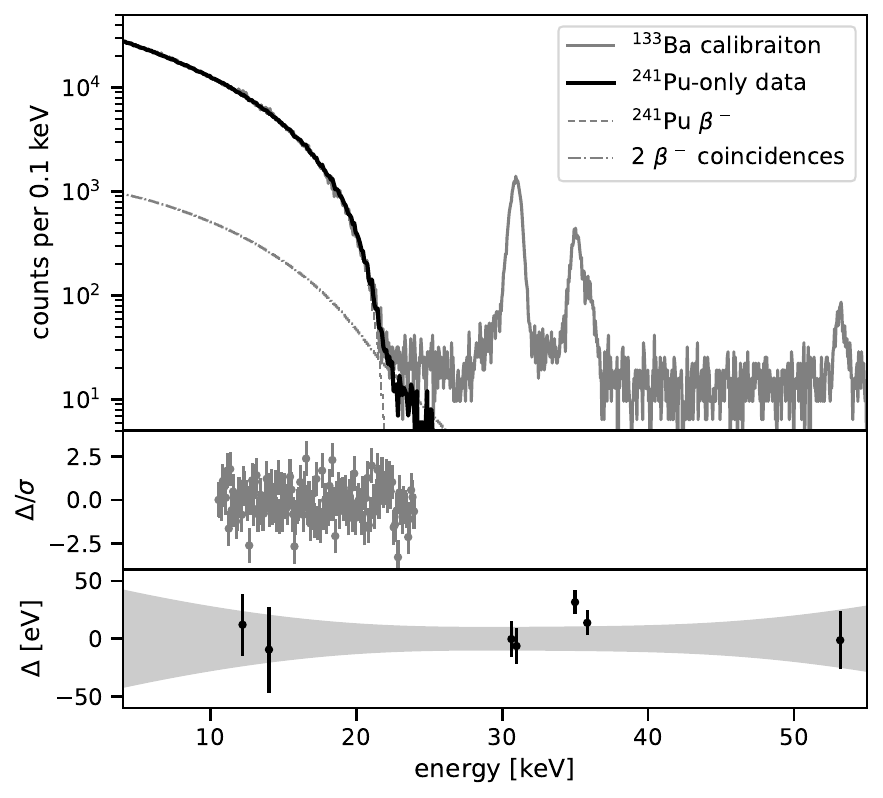}
};
\node[align=left] at (-0.8,3.2) {81-keV $\gamma$  \\- Au $K_\alpha$};
\draw[gray, thick, ->] (-1.5,3.3) -- (-1.9,3.);
\draw[gray, thick, ->] (-1.5,3.1) -- (-1.7,2.9);
\node[] at (0.7,2.3) {Cs $K_\alpha$};
\node[align=right] at (1.7,1.8) {Cs $K_\beta$};
\node[align=right] at (3.3,1.2) {53.2-keV $\gamma$};
\end{tikzpicture}
\caption{\label{fig:Ba133spectra}
(Top) Run 151 spectra with (gray) and without (black) the external $^{133}$Ba source. The source-on spectrum includes X-ray and $\gamma$-ray peaks. The source-off spectrum is fitted with $^{241}$Pu $\beta$-decay model (dashed gray) and its double-coincidence model (dot-dashed gray). (Middle) Residuals of the source-off spectrum. (Bottom) Calibration residuals and uncertainty.}
\end{figure}

The measured data were processed using the same methodology as Run~122, with optimizations for the energy resolution near $Q_\beta$. The sensitivity drift was corrected using the internal $\alpha$ decays. We calibrated the detector by fitting Eq.~(\ref{eq:quadratic}) to the source-on data without the $e_0$ constraint, using a local rather than global calibration to suppress systematic uncertainty from ADC nonlinearity~\cite{stj_nonlinearity}. The resulting calibration was then applied to the source-off data. The top panel of Figure~\ref{fig:Ba133spectra} shows the energy spectra from Run 151, and the bottom panel confirms the accuracy of the quadratic fit via the residuals. The calibration uncertainty at 22\,keV was 12\,eV.

For the $Q_\beta$ analysis, the source-off data were fitted to Eq.~(\ref{eq:beta_decay}) over 10.5–24.5\,keV. 
The fit included a background model for accidental coincidences of two $^{241}$Pu $\beta$ decays, as detailed in Sec.~\ref{subsection:spectral_model}. 
The energy resolution, $\sigma_{151} = $ 0.268\,(24)\,keV, was estimated by analyzing the 12.191\,keV $K_{\alpha1}$ escape peak and the 53.2\,keV $\gamma$ peak. 
The Lorentzian broadening of the $K_{\alpha1}$ peak was 24\,eV and therefore negligible. 
The estimated $\sigma_{151}$ was incorporated into the fit as a Gaussian constraint. 
The best-fit result, obtained by minimizing the extended binned negative log-likelihood (NLL) and shown in Figure~\ref{fig:Ba133spectra}, yielded
\begin{equation}
    Q_\beta = 22.273(33)\,\text{keV}.
\label{eq:Qb}
\end{equation}

This value differs from the accepted 20.78\,(17)\,keV and remains higher than the value reported by Loidl \emph{et al.}~\cite{LOIDL20101454}. Table~\ref{tab:endpoint} summarizes the uncertainty budget for this measurement. Uncertainties were evaluated by repeating the analysis with slightly varied input parameters; for example, $Q_\beta$ shifted by 12\,eV when the atomic-overlap correction factor $B''$ was changed by 5\%. The total variance was calculated as the quadratic sum of the individual contributions. Without the overlap correction, directly comparable to Loidl \emph{et al.}, the result was $Q_\beta = 22.04$\,keV.

\begin{table*}
\caption{\label{tab:endpoint} $^{241}$Pu $Q_\beta$ uncertainty budget}
\begin{ruledtabular}
\begin{tabular}{H{0.18\linewidth} c c L{0.55\linewidth}}
source & $\delta Q_\beta$ (eV) & $\delta Q_\beta/Q_\beta$ (\%) & comment\\
\colrule
Statistical & 15 & 0.067 & Uncertainty from the fitting process.\\
Energy calibration & 8 & 0.036 & From the calibration uncertainty band in Fig.~\ref{fig:Ba133spectra}.\\
Energy resolution & 3 & 0.013 & From the 24\,eV uncertainty of the 12.191\,keV $K_{\alpha1}$ escape and 53.2\,keV $\gamma$ peaks; indirectly included in the total uncertainty through the statistical term.\\
Atomic overlap correction factor $B''$ & 12 & 0.054 & Derived from the 5\% theoretical uncertainty of $B''$.\\
Background & 25 & 0.112 & Change in coincidence background when extending the fit range to 10--27\,keV.\\
$\beta$ coincidence model & 5 & 0.022 & Difference relative to a model using $Q_\beta=21.6$\,keV instead of 22.273\,keV.\\
\colrule
Total & 33 & 0.150 & Combined in quadrature.\\
\end{tabular}
\end{ruledtabular}
\end{table*}

\section{$^{241}$P\MakeLowercase{u} $\beta$ decay spectrum shape \& HNL search}
We used the Run~122 low-energy spectra to measure the $^{241}$Pu $\beta$ decay spectral shape and to search for HNLs. The spectra were calibrated using the $Q_\beta$ value determined in Run~151, and they were compared with theoretical models and the previous measurements of Loidl \emph{et al.}~\cite{LOIDL20101454}. Finally, we constrained the possible admixture of HNLs by comparing the measured spectra with a series of HNL hypotheses.

\subsection{Spectrum model}\label{subsection:spectral_model}
The spectrum model $R_{\text{active}}$ represents the low-energy population as the sum of single-decay events, $R_\beta$, and background contributions, $R_\text{bg}$: 
\begin{align} 
R_\text{active}(E) = n_1 N(E, Q_\beta) + R_\text{bg}(E, n_2, n_3, n_{^{237}\text{U}}). \label{eq:fit_model} \end{align} 
$N(E,Q_\beta)$ follows Eq.~(\ref{eq:beta_decay}), but its amplitude $n_1$ and endpoint energy $Q_\beta$ are free parameters, while $m_\nu$ is fixed to the accepted neutrino masses. $R_\text{bg}$ accounts for the accidental coincidences of two or three $^{241}$Pu $\beta$ decays of amplitude $n_2$ and $n_3$ and a nearly flat component from $^{237}$U $\beta$ decays~\cite{MOUGEOT2023111018}. Finally, $R_\text{active}(E)$ is convolved with a Gaussian response function~\footnote{This approximation of the asymmetric response, similar to that used for $\alpha$ peaks, is justified because the expected energy loss due to self-absorption is two orders of magnitude smaller than the energy resolution at 20\,keV.} with a standard deviation $\sigma$ to account for detector energy resolution.

\subsection{Accidental coincidences}\label{subsection:pileup}
\begin{figure}
\includegraphics[width=0.8\linewidth]{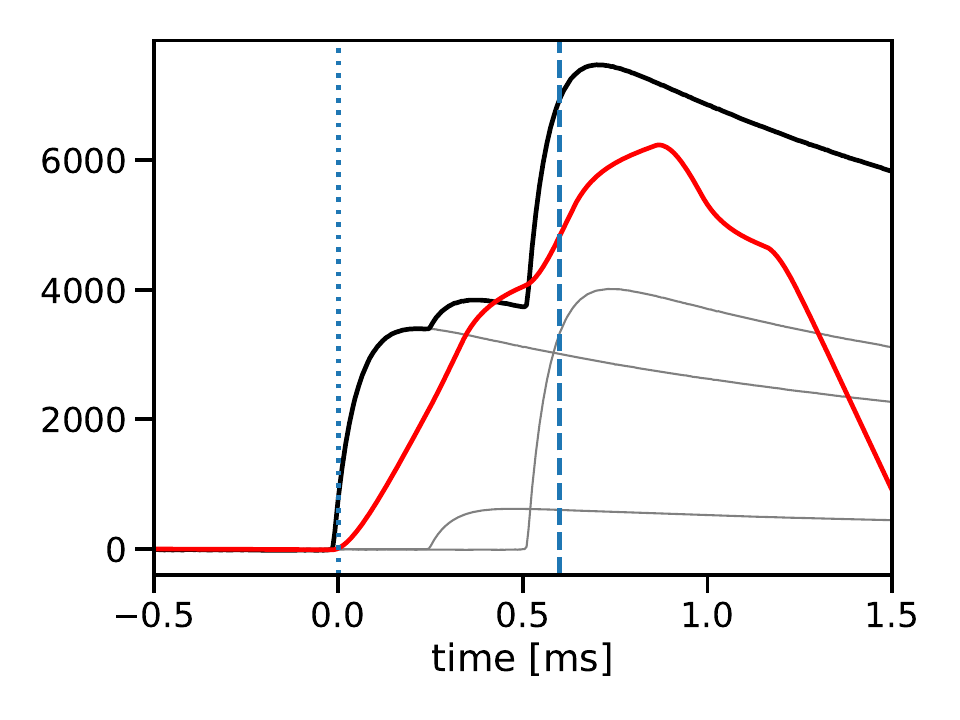}
\caption{\label{fig:pileup}
Accidental-coincidence modeling. Coincidence waveform (black) is the sum of three constituent pulses (gray) with random onset times and amplitudes. This waveform was processed identically to the Run~122 data (see Figure~\ref{fig:processing}), yielding \emph{long pulse} (red). \emph{Pulse-onset} time $t_0$ (dotted) is measured from the first constituent pulse, and the amplitude is extracted at $t_a$ after $t_0$ (dashed).}
\end{figure}

The dominant low-energy background arose from the accidental coincidences in the Run~122 spectra. At the $\sim90$\,cts/s trigger rate and with the processing windows of 1.25 and 1\,ms, the total coincidence probabilities are 15.2\% and 10.3\% for Channel~0 and Channel~1, respectively. Developing analysis cuts for these events was challenging because the cut efficiency significantly decreased at low energies, where the signal-to-noise ratio was reduced.

Instead, we constructed coincidence spectra by reproducing coincidence waveforms using a statistical model and pulse templates. We generated ten million simulated coincidence waveforms, each composed of two or more constituent pulses. The amplitudes of the constituent pulses were randomly sampled from the best-estimate theoretical $\beta$-decay model, and their timestamps were assigned randomly.
The resulting coincidence waveforms were processed in the same manner as the Run~122 data, and the corresponding amplitude distribution defined the accidental-coincidence model used in $R_\text{bg}$. An example coincidence waveform and its corresponding trapezoidal shaping are shown in Figure~\ref{fig:pileup}.

\subsection{Low energy calibration}


The low-energy Run~122 spectra were rescaled so that their $Q_\beta$ values matched those from Run~151. Fitting Eq.~(\ref{eq:fit_model}) to the Run~122 spectra before rescaling yielded $Q_\beta$ values of 22.050\,keV and 21.645\,keV for Channel~0 and Channel~1, respectively. The deviations from the Run~151 result, attributed to higher-order nonlinearities, were modest at 1.0--2.8\%, indicating that Eq.~(\ref{eq:quadratic}) remained a good description of the response. Several features of the data enable such linear rescaling. The fitted $Q_\beta$ values were derived from the full spectral shape and carried small statistical uncertainties of 4\,eV and 5\,eV, providing precise calibration points. The detector response was highly linear over the fitted range, since the quadratic calibration canceled most of the residual nonlinearity. Finally, the calibration simultaneously accounted for the zero-energy offset by incorporating the measured \emph{noise amplitude}.


\subsection{$^{241}$P\MakeLowercase{u} $\beta$ decay spectrum shape}\label{subsection:pu241spectralShape}

\begin{figure*}
\begin{tabular}{cc}
    \centering
    \begin{tikzpicture}
    \node (image) at (0,0) {
    \includegraphics[width=0.5\linewidth]{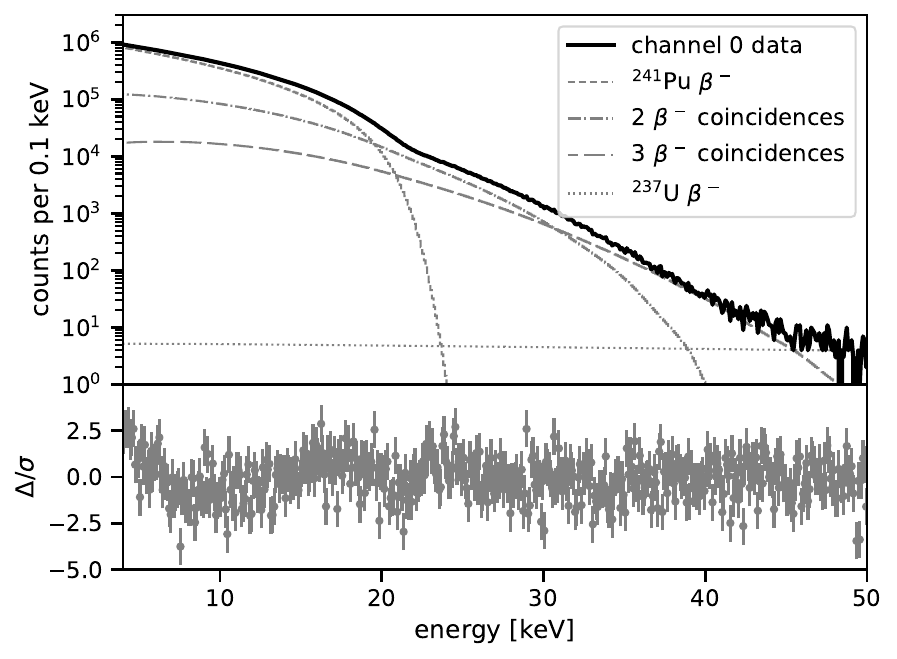}
    };
    \end{tikzpicture}
     & \begin{tikzpicture}
    \node (image) at (0,0) {
    \includegraphics[width=0.5\linewidth]{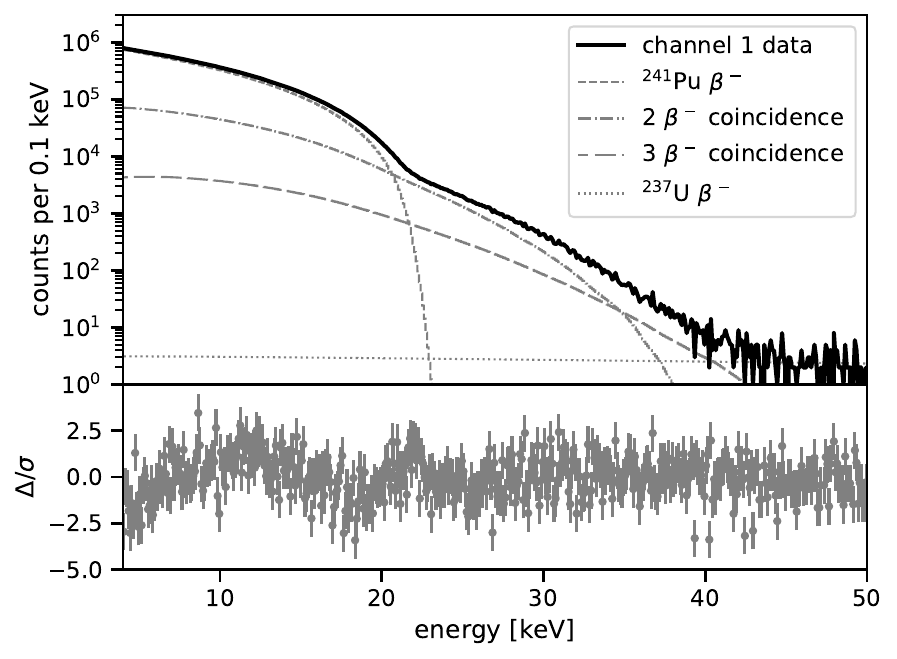}
    };
    \end{tikzpicture}
\end{tabular}
\caption{\label{fig:global_fits}
Fits to the measured data from Channel~0 (left) and Channel~1 (right). Each fit includes contributions from $^{241}$Pu $\beta$ decays, accidental coincidences, and $^{237}$U $\beta$ decays. Bottom panels show the fit residuals, where positive values indicate that the observed data exceed the model prediction.}
\end{figure*}

\begin{figure*}
\begin{tabular}{cc}
    \centering
    \begin{tikzpicture}
    \node (image) at (0,0) {
    \includegraphics[width=0.5\linewidth]{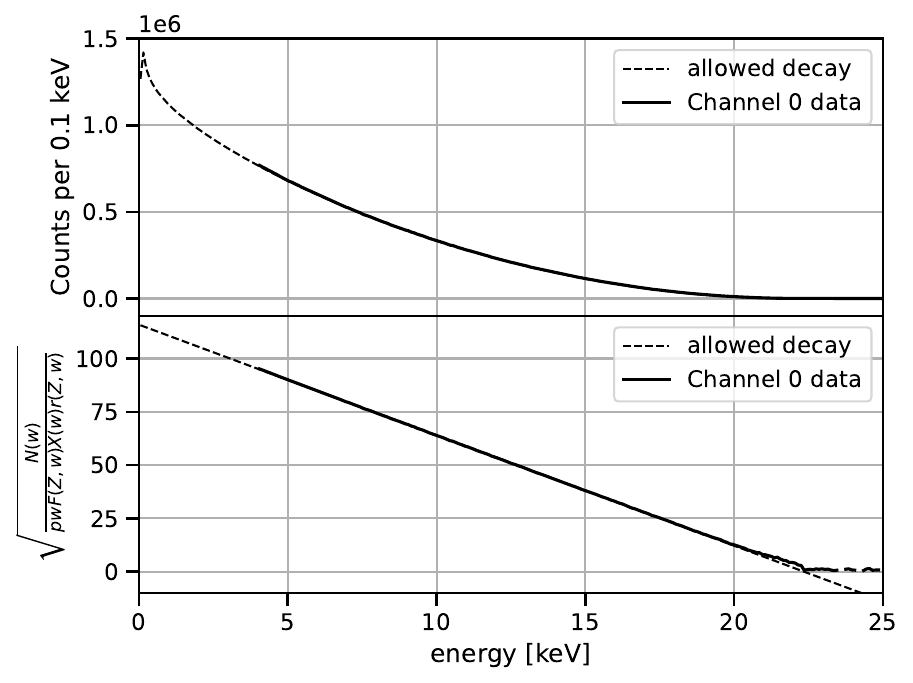}
    };
    \end{tikzpicture}
     & \begin{tikzpicture}
    \node (image) at (0,0) {
    \includegraphics[width=0.5\linewidth]{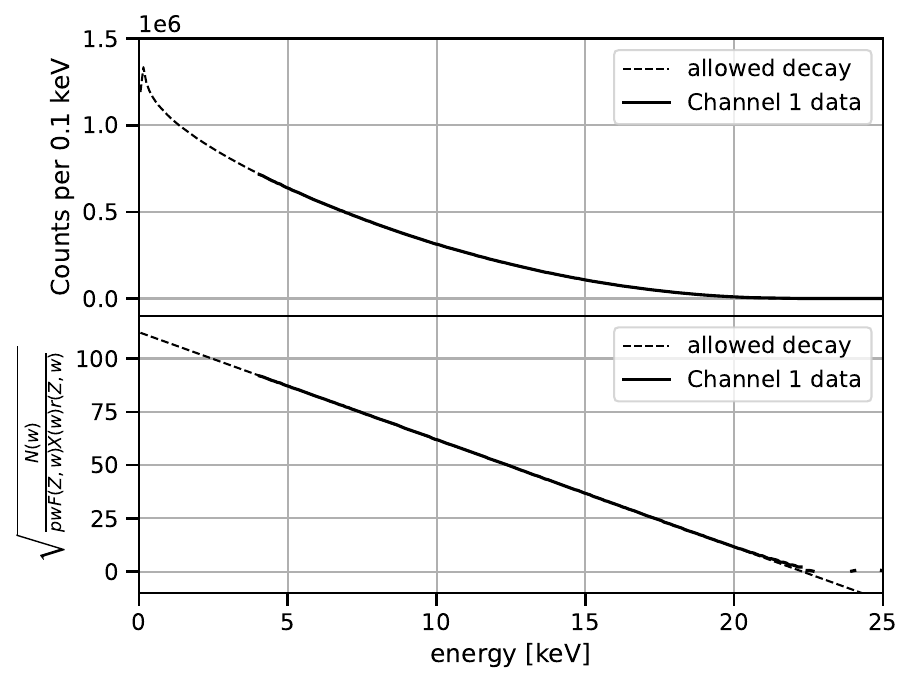}
    };
    \end{tikzpicture}
\end{tabular}
\caption{\label{fig:kurie}
Background-subtracted $^{241}$Pu $\beta$ decay spectra for Channel~0 (left) and Channel~1 (right). Top panels show the $\beta$ differential rate on a linear scale, and bottom panels present the corresponding Kurie plots, in which the allowed decay model appears linear. Spectra from both channels agree closely with the allowed decay model when atomic corrections are included.}
\end{figure*}

We measured the $^{241}$Pu $\beta$ decay spectral shape and performed HNL searches by fitting the rescaled Run~122 spectra to Eq.~(\ref{eq:fit_model}) over the 4--50\,keV range. 
The fits are shown in Figure~\ref{fig:global_fits}, and the corresponding best-fit parameters are summarized in Table~\ref{tab:fit_results}. 
Above 6.5\,keV (Channel~0) and 8.5\,keV (Channel~1), the spectra were statistically consistent ($p\text{-value} > 0.05$) with $R_\text{active}(E)$, and most residuals in the lower panels lay within $3\sigma$. 
The fit quality degraded at lower energies, as indicated by the increasing residuals. Because the residuals from Channels~0 and~1 were uncorrelated, we attributed these discrepancies to systematic effects, such as ADC nonlinearity and temperature-dependent signal-shape variations~\cite{zahir2024energy}, rather than to an energy-dependent shape factor or the presence of HNLs. 
The best-fit $\sigma$ exceeded the baseline resolution, likely due to uncertainties in the sensitivity-drift correction across datasets.

\begin{table}[b]
\caption{\label{tab:fit_results} Run-122 low energy data fit results}
\begin{ruledtabular}
\begin{tabular}{c|cc}
fit parameter & Channel 0 & Channel 1\\
\colrule
$\sigma$ [keV] & 0.736 (14) & 0.531 (15)\\
single $\beta$ & 85.51 (9) $\times 10^6$  &  80.38 (7) $\times 10^6$  \\
two $\beta$ coinc. & 15.81 (12) $\times 10^6 $ &  8.61 (9) $\times 10^6$ \\
three $\beta$ coinc. & 2.95 (4) $\times 10^6$ & 0.65 (3) $\times 10^6$  \\
total  $^{241}$Pu $\beta$ & 104.27 (15) $\times 10^6$ & 89.64 (12) $\times 10^6$ \\
$^{237}$U $\beta$ decays &  2.67 (2) $\times 10^3$ & 1.60 (13) $\times 10^3$ \\
\end{tabular}
\end{ruledtabular}
\end{table}

Figure~\ref{fig:kurie} shows the background-subtracted $^{241}$Pu $\beta$ spectra for clearer visualization, with theoretical allowed-decay spectra overlaid for comparison. The lower panels display the same data as the Kurie plots. The Kurie factors include all atomic corrections, following Eq.~(\ref{eq:Kurie}), so the allowed decays appear linear. The measured spectra again agree closely with the allowed-decay model, showing no significant evidence of an energy-dependent shape factor.

\begin{figure}
\includegraphics[width=\linewidth]{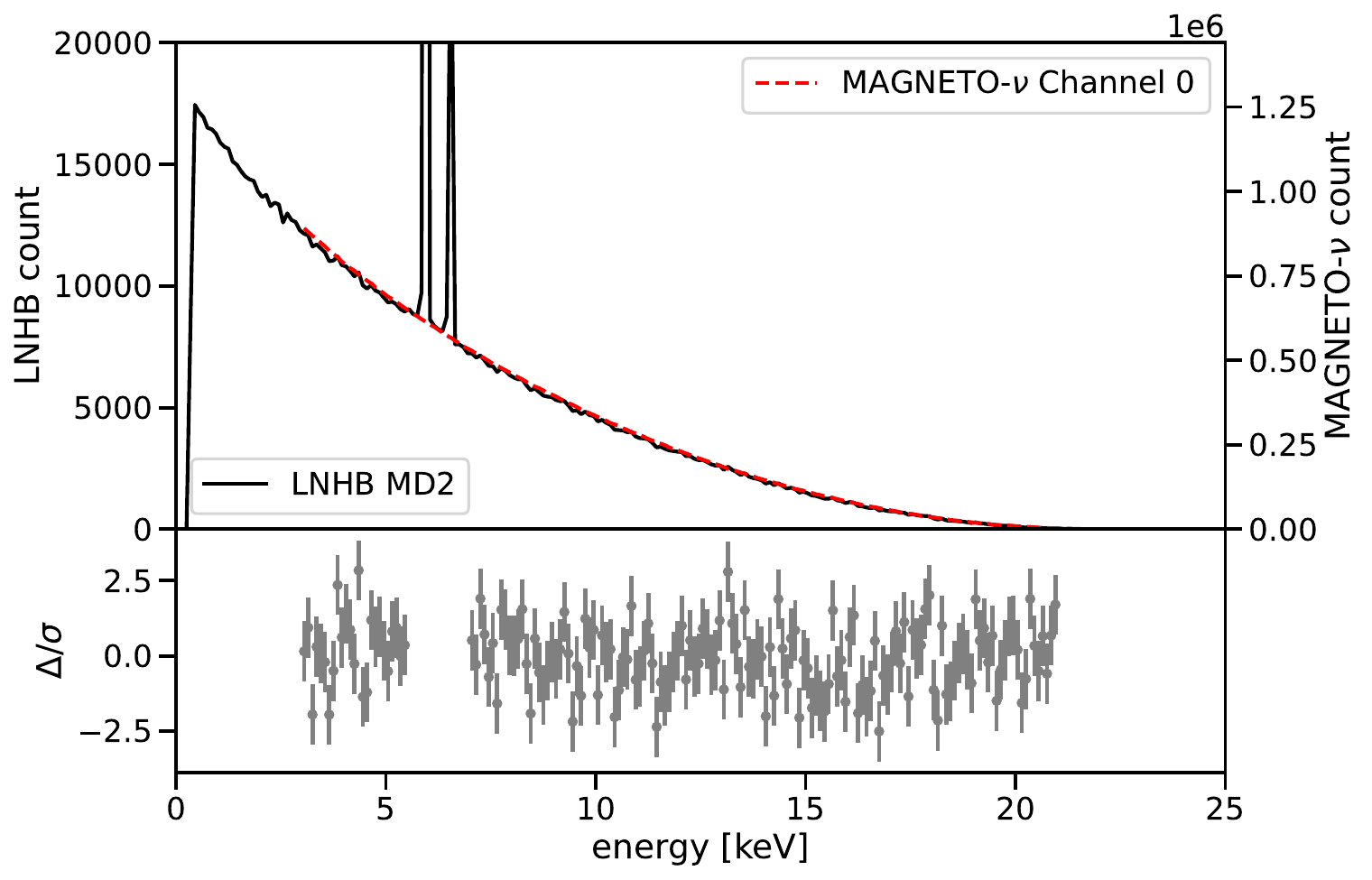}
\caption{\label{fig:lhnb_comp}
Comparison of $^{241}$Pu $\beta$ spectra from Loidl \emph{et~al.}~\cite{LOIDL20101454} (LNHB, black) and from MAGNETO-$\nu$ Channel~0 (red). The Loidl spectrum includes $^{55}$Mn X-ray calibration peaks. The Channel~0 spectrum is plotted on a separate $y$-axis (right) for direct visual comparison. Bottom panel shows the fit residuals.}
\end{figure}

Finally, Figure~\ref{fig:lhnb_comp} compares the Channel~0 spectrum with that of Loidl \emph{et al.} (LNHB)~\cite{LOIDL20101454}. For direct comparison, the LNHB spectrum is linearly scaled with a constant-background component, and the MAGNETO-$\nu$ spectrum is rescaled to match the $Q_\beta$ value. The two spectra show good agreement ($\chi^2$/n.d.f = 1.2).

\subsection{ADC nonlinearity}
The residuals in Figure~\ref{fig:global_fits} are small (on the order of 0.1\%), yet they can still affect the HNL-search sensitivity. A plausible source of these residuals is the nonlinearity and non-uniformity of the ADCs used in the experiment. This effect is well documented in cryogenic-microcalorimeter studies, with multiple prior cases reported~\cite{stj_nonlinearity, stj-nonuniformity, beest_daq, PhysRevD.111.052010}. Uneven ADC binning can distort spectra, particularly when signals occupy only a small fraction of the full input range due to a wide dynamic range. For instance, a 5-keV signal spanned $\approx 0.03$\% of the full input voltage range of the Run~122 setup.

\begin{figure}
\begin{tikzpicture}
\node (image) at (0,0) {
    \includegraphics[width=\linewidth]{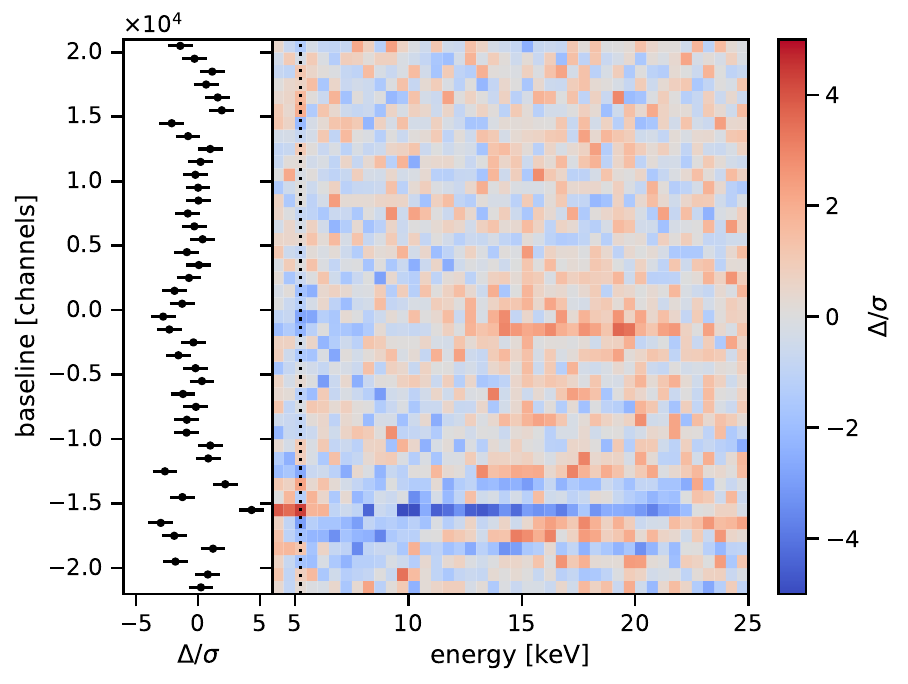}
    };
\draw[thick, ->] (.5,-0.8) -- (0.5,-1.5);

\end{tikzpicture}
\caption{\label{fig:ch0_residual_base}
Residuals of Channel~0 data as a function of energy and baseline. 
Left panel shows the variation of residuals at different baselines for 5.25\,keV (vertical dotted in the main panel).}
    \label{fig:ch0_residual_base}
\end{figure}

Figure~\ref{fig:ch0_residual_base} reveals a potential correlation between the observed residuals and the signal baseline. It shows the residuals of Channel 0 data across energy and baseline. In particular, data near ADC channel -15000 counts (black arrow) exhibit a distinct residual pattern compared to the rest of the distribution, indicating stronger ADC nonlinearity in this range. Because most events fell near this baseline, these residuals dominated the overall pattern. The high statistical precision of the data further accentuated these residuals. Future work will focus on characterizing and correcting ADC nonlinearity~\cite{9288848} and refining the data-acquisition scheme to mitigate its impact. The PXIe-6356 digitizer, used for the remaining MAGNETO-$\nu$ data, is reported to exhibit a more linear response~\cite{beest_daq}.

\subsection{HNL search}
To search for HNLs, we compared the test statistic of each HNL hypothesis with that of the null hypothesis ($|U_{e4}|^2 = 0$). 
Our signal model incorporates the effects of ADC nonlinearity, which can distort the spectral shape but preserves the total event count. 
A quadratic function with three free parameters provides the simplest such representation:
\begin{equation}
    \lambda_\text{ANL} = j(E-h)^2 + k,
\end{equation}
where $h$, $j$, and $k$ are free fit parameters. 
These nuisance parameters improve the model fit but reduce the overall HNL-search sensitivity. 
The signal model is defined as the normalized product of $R_\beta$ [Eq.~(\ref{eq:sterile_rate})] and $\lambda_\text{ANL}$:
\begin{equation}
    \tilde{R_\beta}(E, m_4^2, |U_{e4}|^2) = \text{Norm}(R_\beta \times \lambda_\text{ANL}).
\end{equation}
The sum of $\tilde{R_\beta}(E, m_4^2, |U_{e4}|^2)$ and $R_\text{bg}$ forms the total expected spectral model:
\begin{align}
    R_\text{HNL}(E, m_4^2, |U_{e4}|^2) &= n_1 \tilde{R_\beta}(E, m_4^2, |U_{e4}|^2) \nonumber \\
    &+ R_\text{bg}(E, n_2, n_3, n_{^{237}\text{U}}). \label{eq:hnl_search}
\end{align}
To keep the $\lambda_\text{ANL}$ model simple and computationally efficient, we approximate that $\lambda_\text{ANL}$ does not apply to $R_\text{bg}$, which is a subdominant component at low energies. 
Finally, $R_\text{HNL}$ is convolved with a Gaussian detector-response function of width $\sigma$.

The likelihood minimizer compared Eq.~(\ref{eq:hnl_search}) to the low-energy Run 122 spectra as discussed in Section~\ref{subsection:pu241spectralShape} and searched for the lowest NLL by varying all nuisance parameters except $m_4$ and $|U_{e4}|^2$. For each HNL scenario defined on a 2D grid spanning (0.5, 19.5)\,keV for $m_4$ and ($10^{-4}$, 0.1) for $|U_{e4}|^2$, the fitting process was repeated, and NLL was collected. The test statistic $\Delta\mathcal{L}$ was defined as the difference between the NLL of a given HNL scenario and that of the null hypothesis, $|U_{e4}|^2=0$. 

\begin{figure}
\includegraphics[width=\linewidth]{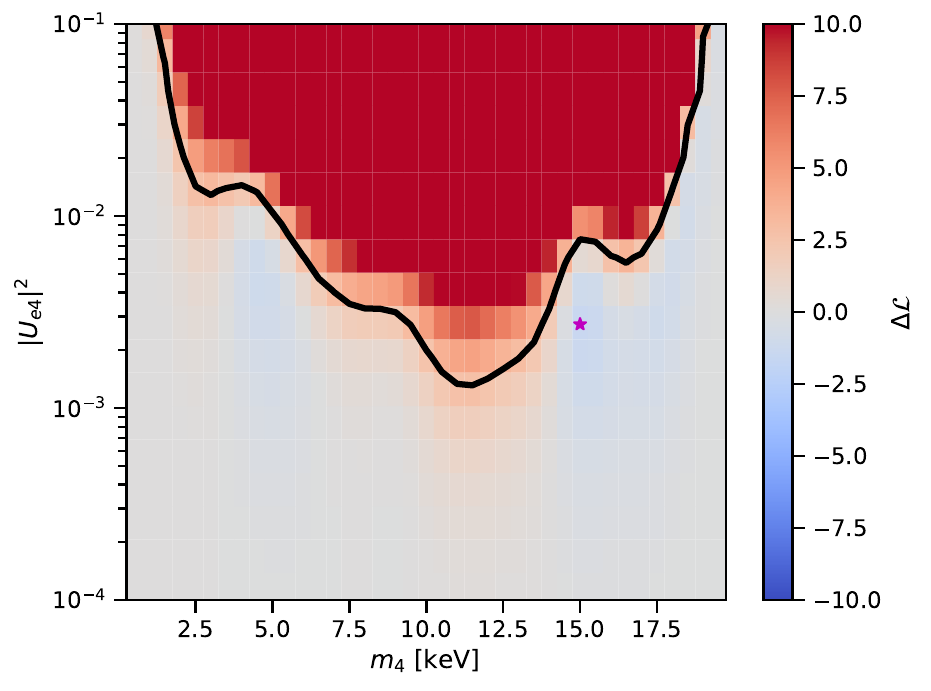}
\includegraphics[width=\linewidth]{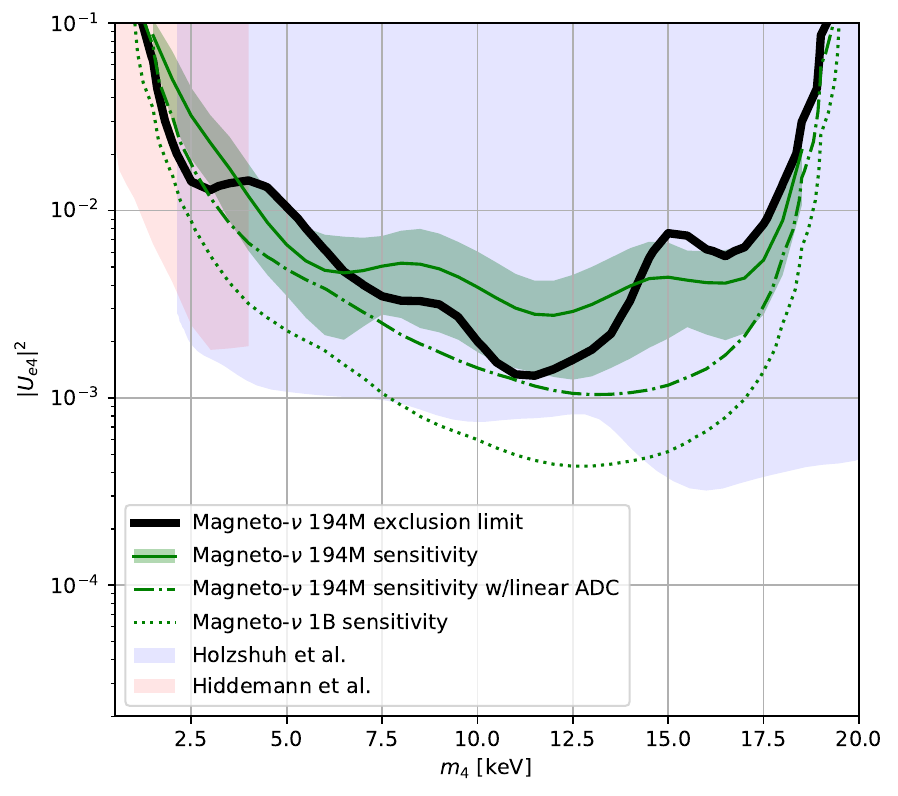}
\caption{\label{fig:nll_map}
(Top) Test statistic $\Delta\mathcal{L}$ from Run~122 as a function of HNL mass $m_4$ and admixture $|U_{e4}|^2$. Region above the solid black contour is excluded at the 95\% confidence level. Magenta star marks the best-fit point. (Bottom) Sensitivity (green line and band) and exclusion limit (solid black) derived from the initial 194 million decay events of the first MAGNETO-$\nu$ dataset. Blue and red shaded regions are excluded by $^{63}$Ni~\cite{HOLZSCHUH1999247} and tritium~\cite{Hiddemann_1995} $\beta$-decay measurements, respectively.}
\end{figure}

The top plot of Figure~\ref{fig:nll_map} shows the $\Delta\mathcal{L}$ values summed from both channels. According to Wilks' theorem, the $\Delta\mathcal{L}$ distribution is expected to follow one-half of a chi-squared distribution with two degrees of freedom, corresponding to the independent choices of $m_4^2$ and $|U_{e4}|^2$. For this distribution,  the 95\% confidence level corresponds to $\Delta\mathcal{L}=2.996$. Consequently, parameter space exceeding this threshold, outlined by the black contour line, was excluded at the 95\% confidence level. The lowest exclusion limit achieved in our analysis was $|U_{e4}|^2 = 1.31 \times 10^{-3}$ at $m_4 = 11.5$\,keV. The best fit point of $R_\text{HNL}$ occurred at $m_4 = 15$\,keV and $|U_{e4}|^2 = 2.7 \times 10^{-3}$. Because its significance (1.41) falls below the 95\% threshold, only exclusion limits are presented.

To estimate the Phase~I sensitivity of MAGNETO-$\nu$, we recalculated $\Delta\mathcal{L}$ using toy Monte Carlo datasets generated from the $\beta$-decay model. Averaging the $\Delta\mathcal{L}$ values from 30 toy datasets reduced statistical fluctuations, and the resulting mean and standard deviation appear as the green line and band in Figure~\ref{fig:nll_map}. The figure also demonstrates that improved characterization of ADC nonlinearity substantially increases sensitivity (dash-dotted line). Increasing the dataset to one billion decays is projected to enhance the sensitivity further (dotted line), exceeding the limit set by Holzschuh \emph{et al.} around 12.5\,keV.

\section{Conclusion}
We presented the MAGNETO-$\nu$ experiment, which searched for heavy neutral leptons (HNLs) in the $\mathcal{O}(10)$\,keV mass range using high-precision $^{241}$Pu $\beta$ spectra measured with an MMC-based DES spectrometer. The experiment collected the highest-statistics $^{241}$Pu $\beta$ spectrum to date, accompanied by a rich set of actinide $\alpha$-decay data used for energy calibration and stability monitoring.

Using an external $^{133}$Ba calibration source, we obtained $^{241}$Pu $Q_\beta$ = 22.273\,(33)\,keV, which is essential for modeling the beta spectrum. This value is slightly higher than that reported by Loidl \emph{et al.} and substantially higher than the accepted literature value of 20.78\,(17)\,keV. The measured spectral shape agrees well with the theoretical allowed-decay model once atomic effects are included. Small residual deviations from the model at low energies are attributed to systematic effects.

From the initial sample of 194 million $\beta$-decay events, we established the most stringent HNL-mixing limit yet derived from the $^{241}$Pu spectrum, comparable to the best existing constraints. This exclusion limit will be further improved if the aforementioned systematic effects are better characterized and controlled. We estimate that analysis of one billion events will reach a sensitivity of $|U_{e4}| \approx 4 \times 10^{-4}$ for an HNL mass around 13\,keV, surpassing the existing limit set by Holzschuh \emph{et al.}


\begin{acknowledgments}
This work was performed under the auspices of the U.S. Department of Energy by Lawrence Livermore National Laboratory under Contract DE-AC52-07NA27344. This work was funded by the Laboratory Directed Research and Development program of Lawrence Livermore National Laboratory (20-LW-024 and 23-LW-043). This work was funded by the National Nuclear Security Administration of the Department of Energy, Office of Defense Nuclear Nonproliferation Research and Development (NA-22) and International Nuclear Safeguards (NA-241). This material is based upon work supported by the Consortium for Nuclear Forensics under Department of Energy, National Nuclear Security Administration award number DE-NA0004142.
\end{acknowledgments}

\bibliography{magneto2}
\bibliographystyle{apsrev4-2-trunc15}
\end{document}